\documentclass[amsmath,amssymb, aps, prb, twocolumn, notitlepage]{revtex4-2}
\usepackage[utf8]{inputenc}
\usepackage{braket}
\usepackage{indentfirst}
\usepackage{graphicx}
\usepackage{dcolumn}
\usepackage{bm}
\usepackage[usenames]{color}
\usepackage{slashed}
\usepackage[utf8]{inputenc}
\usepackage{amsfonts}
\usepackage{amsmath}
\usepackage{amssymb}
\usepackage{hyperref}
\usepackage{latexsym}
\usepackage{color}
\usepackage{soul}
\usepackage{appendix}

\begin{document}
\title{Maximum-Likelihood-Estimate Hamiltonian learning via efficient and robust quantum likelihood gradient} 
\author{Tian-Lun Zhao}
\affiliation{International Center for Quantum Materials, School of Physics, Peking University, Beijing, 100871, China}
\author{Shi-Xin Hu}
\affiliation{International Center for Quantum Materials, School of Physics, Peking University, Beijing, 100871, China}
\author{Yi Zhang}
\email{frankzhangyi@gmail.com}
\affiliation{International Center for Quantum Materials, School of Physics, Peking University, Beijing, 100871, China}

\date{Today}

\begin{abstract}
Given the recent developments in quantum techniques, modeling the physical Hamiltonian of a target quantum many-body system is becoming an increasingly practical and vital research direction. Here, we propose an efficient strategy combining maximum likelihood estimation, gradient descent, and quantum many-body algorithms. Given the measurement outcomes, we optimize the target model Hamiltonian and density operator via a series of descents along the quantum likelihood gradient, which we prove is negative semi-definite with respect to the negative-log-likelihood function. In addition to such optimization efficiency, our maximum-likelihood-estimate Hamiltonian learning respects the locality of a given quantum system, therefore, extends readily to larger systems with available quantum many-body algorithms. Compared with previous approaches, it also exhibits better accuracy and overall stability toward noises, fluctuations, and temperature ranges, which we demonstrate with various examples. 
\end{abstract}

\maketitle

\section{Introduction}

Understanding the quantum states and the corresponding properties of a given quantum Hamiltonian is a crucial problem in quantum physics. Many powerful numerical and theoretical tools have been developed for such purposes and made compelling progress \cite{Lanczos, SteveWhite1992, dmrg, QMC, rgf}. On the other hand, with the rapid experimental developments of quantum technology, e.g., near-term quantum computation \cite{nielsen2002, Nayak2008} and simulation \cite{Buluta2009, Georgescu2014, Barthelemy2013, Browaeys2020, Scholl2021, Ebadi2022, Bluvstein2021}, it is also vital to explore the inverse problem, e.g., Hamiltonian learning - optimize a model Hamiltonian characterizing a quantum system with respect to the measurement results. Given the knowledge and assumption of a target system, researchers have achieved many resounding successes modeling quantum Hamiltonians with physical pictures and phenomenological approaches \cite{tbg, bcs}. However, such subjective perspectives may risk biases and are commonly insufficient on detailed quantum devices. Therefore, the explorations for objective Hamiltonian learning strategies have attracted much recent attention \cite{Qi2019, FETH, Corr1, RN571, Corr2, Entangle, dynamic1, Wenjunyu, Hsinyuan}.

There are mainly two categories of Hamiltonian-learning strategies, based upon either quantum measurements on a large number of (identical copies of) quantum states, e.g., Gibbs states or eigenstates \cite{Qi2019, FETH, Corr1, RN571, Corr2, Entangle}, or initial states' time evolution dynamics \cite{dynamic1, Wenjunyu, Hsinyuan, TNmethod}, corresponding to the target quantum system. For example, given the measurements of the correlations of a set of local operators, the kernel of the resulting correlation matrix offers a candidate model Hamiltonian \cite{Qi2019, FETH, Corr1}. On the other hand, while established theoretically, most approaches suffer from elevated costs and are limited to small systems in experiments or numerical simulations \cite{Corr1, RN571, RN572, Zoller}. Besides, there remains much room for improvements in stability towards noises and temperature ranges. 

Maximum likelihood estimation (MLE) is a powerful tool that parameterizes and then optimizes the probability distribution of a statistical model so that the given observed data is most probable. MLE's intuitive and flexible logic makes it a prevailing method for statistical inference. Adding to its wide range of applications, MLE has been applied successfully to quantum state tomography\cite{QSE, PhysRevA.75.042108, Lvovsky_2004, PhysRevA.85.042317,PhysRevA.63.020101}, providing the most probable quantum states given the measurement outputs. 

Inspired by MLE's successes in quantum problems, we propose a general MLE Hamiltonian learning protocol: given finite-temperature measurements of the target quantum system in thermal equilibrium, we optimize the model Hamiltonian towards the MLE step-by-step via a ``quantum likelihood gradient". We show that such quantum likelihood gradient, acting collectively on all presenting operators, is negative semi-definite with respect to the negative-log-likelihood function and thus provides efficient optimization. In addition, our strategy may take advantage of the locality of the quantum system, therefore allowing us to extend studies to larger quantum systems with tailored quantum many-body ansatzes such as Lanczos, quantum Monte Carlo (QMC), density matrix renormalization group (DMRG), and finite temperature tensor network (FTTN) \cite{FTTN, ITensor} algorithms in suitable scenarios. We also demonstrate that MLE Hamiltonian learning is more accurate, less restrictive, and more robust against noises and broader temperature ranges. Further, we generalize our protocol to measurements on pure states, such as the target quantum systems' ground states or quantum chaotic eigenstates. Therefore, MLE Hamiltonian learning enriches our arsenal for cutting-edge research and applications of quantum devices and experiments, such as quantum computation, quantum simulation, and quantum Boltzmann machines \cite{QBM}. 

We organize the rest of the paper as follows: In Sec. II, we review the MLE context and introduce the MLE Hamiltonian learning protocol; especially, we show explicitly that the corresponding quantum likelihood gradient leads to a negative semi-definite change to the negative-log-likelihood function. Via various examples in Sec. III, we demonstrate our protocol's capability, especially its robustness against noises and temperature ranges. We generalize the protocol to quantum measurements of pure states in Sec. IV and Appendix D, with consistent results for exotic quantum systems such as quantum critical and topological models. We summarize our studies in Sec. V with a conclusion on our protocol's advantages (and limitations), potential applications, and future outlooks.

\section{Maximum-likelihood-estimate Hamiltonian learning}

To start, we consider an unknown target quantum system $\hat H_s = \sum_{j} \mu_{j} \hat{O}_{j}$ in thermal equilibrium, and measurements of a set of observables $\{\hat{O}_{i}\}$ on its Gibbs state $\hat{\rho}_{s}=\exp(-\beta\hat{H}_s)/\mbox{tr}[\exp(-\beta\hat{H}_s)]$, where $\beta$ is the inverse temperature. Given a sufficient number $N_i$ of measurements of the operator $\hat{O}_{i}$, the occurrence time $f_{\lambda_i}$ of the $\lambda_i^{th}$ eigenvalue $o_{\lambda_i}$ approaches:
\begin{equation}
f_{\lambda_i}=
p_{\lambda_{i}}N_i\approx\mbox{tr}[\hat{\rho}_{s}\hat{P}_{\lambda_{i}}] N_i,
\end{equation}
where $p_{\lambda_i}=f_{\lambda_i}/N_i$ denotes the statistics of the outcome $o_{\lambda_i}$, and $\hat{P}_{\lambda_i}$ is the corresponding projection operator to the $o_{\lambda_i}$ sector. Our goal is to locate the model Hamiltonian $\hat H_s$ for the quantum system, which commonly requires the presence of all $\hat{H}_s$'s terms in the measurement set $\{\hat{O}_i\}$.

Following previous MLE analysis \cite{QSE, PhysRevA.75.042108, Lvovsky_2004, PhysRevA.85.042317,PhysRevA.63.020101}, the statistical weight of any given state $\hat \rho$ is:
\begin{equation}
    \mathcal{L}(\hat{\rho})\propto \prod_{i, \lambda_i}\{ \mbox{tr}[\hat{\rho}\hat{P}_{\lambda_i}]^{\frac{f_{\lambda_i}}{N_{tot}}}\}^{N_{tot}},  
    \label{LHF}
\end{equation}
upto a trivial factor, where $N_{tot}=\sum_{i}N_i$ is the total number of measurements. For Hamiltonian learning, we search for (the set of parameters ${\mu_j}$ of) the MLE Hamiltonian $\hat{H}$, whose Gibbs state $\hat \rho$ maximizes the likelihood function in Eq. \ref{LHF}. The maximum condition for Eq. \ref{LHF} can be re-expressed as:
\begin{eqnarray}
   &&\hat{R}(\hat{\rho}) \hat{\rho} = \hat{\rho}, \nonumber \\
   &&\hat{R}(\hat{\rho}) = \sum_{i,\lambda_i}\frac{f_{\lambda_i}}{N_{tot}}  \frac{\hat{P}_{\lambda_i}}{\mbox{tr} [\hat \rho \hat P_{\lambda_i}]}, \label{eq:mlerho}
\end{eqnarray}
see Appendix A for a detailed review. Solving Eq. \ref{eq:mlerho} is a nonlinear and nontrivial problem, for which many algorithms have been proposed \cite{ Lvovsky_2004, PhysRevA.75.042108, PhysRevA.85.042317,PhysRevA.63.020101}. For example, we can employ iterative updates $\hat{\rho}_{k+1} \propto \hat{R}(\hat{\rho}_k)\hat{\rho}_k\hat{R}(\hat{\rho}_k)$ until Eq. \ref{eq:mlerho} is fulfilled \cite{Lvovsky_2004}. These algorithms mostly center around the parameterization and optimization of a quantum state $\hat \rho$, whose cost is exponential in the system size. Besides, such iterative updates do not guarantee that the quantum state $\hat{\rho}$ remains a Gibbs form, especially when the measurements are insufficient to uniquely determine the state (e.g., large noises or small numbers of measurements and there are many quantum states satisfying Eq. \ref{eq:mlerho}). Consequently, extracting $\hat{H}\propto-\frac{1}{\beta}\ln\hat{\rho}$ from $\hat \rho$ further adds up to the inconvenience.

\begin{figure}
    \centering
    \includegraphics[width = 0.98\linewidth]{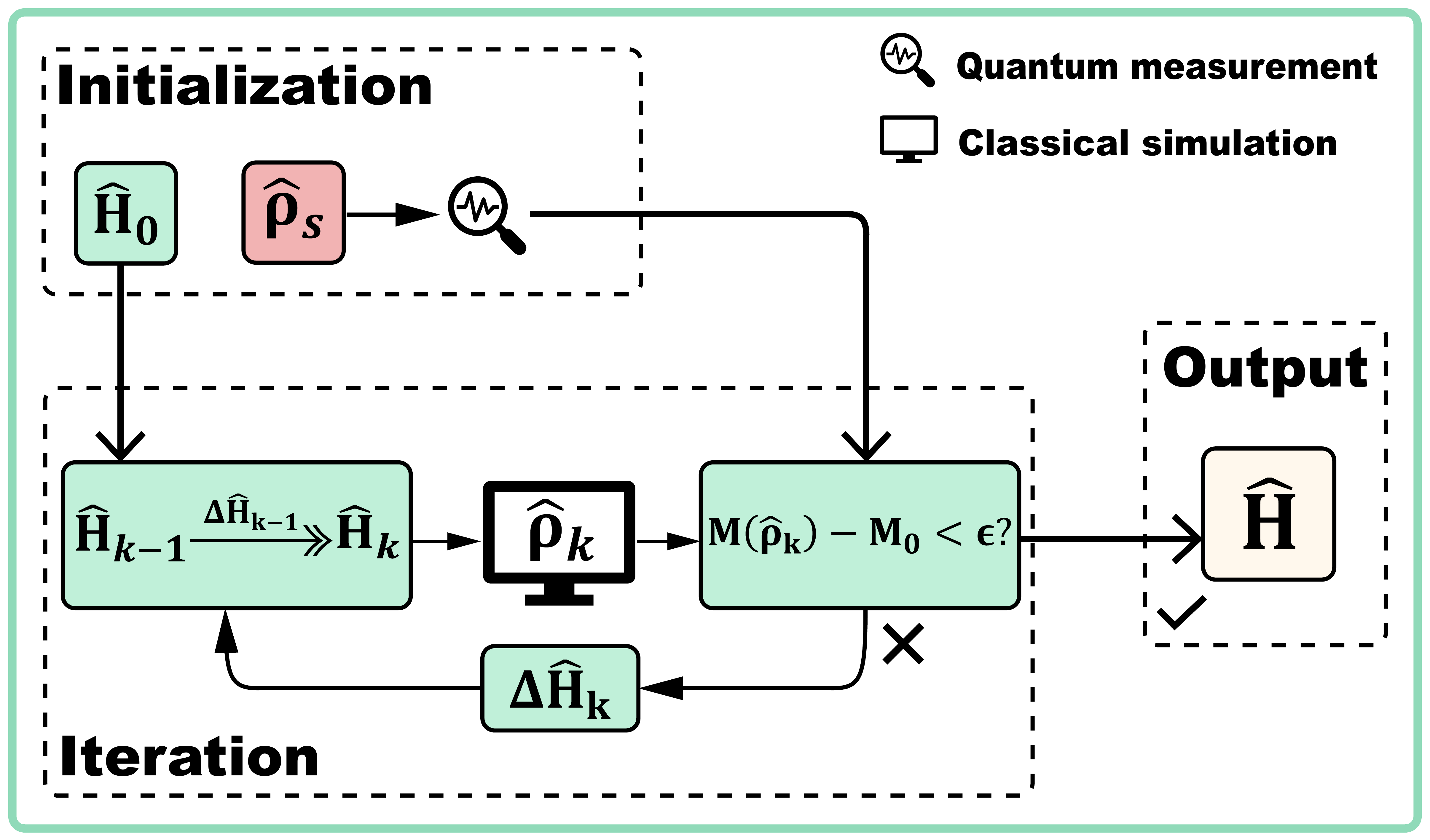}
    \caption{An illustration of the MLE Hamiltonian learning algorithm: given the quantum measurements on the Gibbs state $\hat{\rho}_s$  of the target quantum system, we update the candidate Hamiltonian iteratively until the negative-log-likelihood function (or relative entropy) converges below a given threshold $\epsilon$, after which the output yields the MLE Hamiltonian. Within each iterative step, we evaluate the operator expectation values with respect to the Gibbs state $\hat{\rho}_k = \exp(-\beta\hat H_k) / \mbox{tr}[\exp(-\beta\hat H_k)]$, which directs the model update $\Delta \hat{H}_k=-\gamma\hat{R}_k$ for the next iterative step.}
    \label{fig:algorithm}
\end{figure}

Considering that the operator $\hat{R}(\hat{\rho})$ has the same operator structure as the Hamiltonian, we take an alternative stance for the Hamiltonian learning task and update the candidate Hamiltonian $\hat{H}_k$, i.e., the model parameters, collectively and iteratively. In particular, we integrate the corrections to the Hamiltonian coefficients to the operator $\hat{R}(\hat{\rho})$, which offers such a quantum likelihood gradient (Fig. \ref{fig:algorithm}):
\begin{eqnarray}
    \hat H_{k+1} &=& \hat H_k -\gamma \hat R_k, \nonumber\\
    \hat{\rho}_{k+1}&=&\frac{e^{-\beta\hat{H}_{k+1}}}{\mbox{tr}[e^{-\beta\hat{H}_{k+1}}]}=\frac{e^{-\beta(\hat{H}_{k}-\gamma\hat{R}_{k})}}{\mbox{tr}[e^{-\beta(\hat{H}_{k}-\gamma\hat{R}_{k})}]}, \label{eq:iter}
\end{eqnarray}
where $\gamma>0$ is the learning rate - a small parameter controlling the step size. We denote $\hat{R}_k \equiv \hat{R}(\hat{\rho}_k)$ for short here afterwards. Compared with previous Hamiltonian extractions from MLE quantum state tomography, the update in Eq. \ref{eq:iter} possesses several advantages in Hamiltonian learning. First, we can utilize the Hamiltonian structure (e.g., locality) to choose suitable numerical tools (e.g., QMC and FTTN) and even calculate within the subregions - we circumvent the costly parametrization of the quantum state $\hat{\rho}$. Also, the update guarantees a state in its Gibbs form. Last but not least, we will show that for $\gamma \ll 1$, such a quantum likelihood gradient in Eq. \ref{eq:iter} yields a negative semi-definite contribution to the negative-log-likelihood function, guaranteeing the MLE Hamiltonian (upto a trivial constant) at its convergence and an efficient optimization toward it. 

\textbf{Theorem:} For $\gamma\ll 1$, $\gamma>0$, the quantum likelihood gradient in Eq. \ref{eq:iter} yields a negative semi-definite contribution to the negative-log-likelihood function $ M(\hat{\rho}_{k+1})=-\frac{1}{N_{tot}}\log\mathcal{L}(\hat{\rho}_{k+1})$.

\textbf{Proof:} We note that upto linear order in $\gamma \ll 1$:
\begin{eqnarray}
        e^{-\beta\hat{H}_{k+1}}&=&e^{-\beta\hat{H}_k}\prod_{n=0}^{\infty} \exp\left[\frac{(-1)^{n}}{(n+1)!}{\rm ad}_{-\beta\hat{H}_k}^{n}(\beta\gamma\hat{R}_k)+o(\gamma^2)\right] \nonumber\\
        &\approx& e^{-\beta\hat{H}_k}\left[1+\sum_{n=0}^{\infty}\frac{(-1)^{n}}{(n+1)!}{\rm ad}_{-\beta\hat{H}_k}^{n}(\beta\gamma\hat{R}_k)\right]\nonumber\\
        &=&e^{-\beta\hat{H}_k}(1+\beta\gamma \int_{0}^{1}e^{\beta s\hat{H}_k}\hat{R}_k e^{-\beta s\hat{H}_k}{\rm d}s), \label{eq:linear_gamma}
\end{eqnarray}
where ${\rm ad}_{\hat{A}}^{j}\hat{B}=[\hat{A},{\rm ad}_{\hat{A}}^{j-1}\hat{B}]$ and ${\rm ad}_{\hat{A}}^{0}\hat{B}=\hat{B}$ are the adjoint action of the Lie algebra. The first and third lines are based on the Zassenhaus formula\cite{zassen} and the Baker-Hausdorff formula \cite{zassen}, respectively, while the second line neglects terms above the linear order of $\gamma$. 

Following this, we can re-express the quantum state in Eq. \ref{eq:iter} as:
\begin{equation}
    \hat{\rho}_{k+1}=\hat{\rho}_k\frac{1+\beta\gamma \int_{0}^{1}e^{\beta s\hat{H}_k}\hat{R}_k e^{-\beta s\hat{H}_k}{\rm d}s}{1+\beta\gamma},
    \label{appr_iter}
\end{equation}
where we have used $\mbox{tr}[\hat{\rho}_k \hat{R}_k]=1$ as a direct consequence of $R_k$'s definition in Eq. \ref{eq:mlerho}. 

Subsequently, after introducing the quantum likelihood gradient, the negative-log-likelihood function becomes:
\begin{eqnarray}
   M(\hat{\rho}_{k+1})&=&-\frac{1}{N_{tot}}\log\mathcal{L}(\hat{\rho}_{k+1}) \nonumber\\
   &=&-\sum_{i,\lambda_i}\frac{f_{\lambda_i}}{N_{tot}}\log \mbox{tr}[\hat{\rho}_{k+1}\hat{P}_{\lambda_i}] \nonumber\\
   &\approx& M(\hat{\rho}_k)+\beta\gamma(1-\Delta_k), \label{MLE_iter}
\end{eqnarray}
where we keep terms upto linear order of $\gamma$ in the $\log$ expansion. 

On the other hand, we can establish the following inequality:
\begin{eqnarray}
    \Delta_k &=& \mbox{tr}[\hat{\rho}_k\int_{0}^{1}e^{\beta s\hat{H}_k}\hat{R}_k e^{-\beta s\hat{H}_k}\hat{R}_k{\rm d}s] \nonumber\\
    &=&\int_{0}^{1}\mbox{tr}[\hat{\rho}_k e^{\beta s\hat{H}_k}\hat{R}_k e^{-\beta s\hat{H}_k}\hat{R}_k]\mbox{tr}[\hat{\rho}_k]{\rm d}s \nonumber\\
    &=&\int_{0}^{1}||e^{-\beta s\hat{H}_{k}/2}\hat{R}_{k}e^{-\beta(1-s)\hat{H}_{k}/2}||_{F}^{2}||e^{-\beta\hat{H}_{k}/2}||_{F}^{2}\frac{{\rm d}s}{Z_{k}^2} \nonumber\\
    &\ge& \int_{0}^{1}\mbox{tr}[\hat{\rho}_{k}\hat{R}_{k}]^2{\rm d}s=1, \label{eq:deltagt1}
\end{eqnarray}
where $Z_{k}=\mbox{tr}[e^{-\beta\hat{H}_{k}}]$ is the partition function, $||A||_{F}=\sqrt{\mbox{tr}[A^{\dagger}A]}$ is the Frobenius norm of matrix $A$, and the non-negative definiteness of $\hat \rho_k$ allows $\hat{\rho}_{k}=(\hat{\rho}_{k}^{1/2})^2=(e^{-\beta\hat{H}_{k}/2})^2/Z_{k}$. The inequality in the fourth line follows the Cauchy-Schwarz inequality. 

We note that the equality - the convergence criteria of our MLE Hamiltonian learning protocol - is established if and only if:
\begin{equation}
    e^{-\beta s\hat{H}_{k}/2}\hat{R}_{k}e^{-\beta(1-s)\hat{H}_{k}/2}=e^{-\beta\hat{H}_{k}/2},
\end{equation}
which implies the conventional MLE optimization target $\hat{R}\hat{\rho}=\hat{\rho}$ in Eq. \ref{eq:mlerho}. We can also establish such consistency from our iterative convergence \footnote{In practice, given sufficient measurements, we have $\hat{R}_k\sim \hat{I}$ dictating the quantum likelihood gradient at the iteration's convergence.} following Eq. \ref{eq:iter}:
\begin{equation}
   \hat \rho_{k+1} =\frac{e^{-\beta(\hat{H}_k-\gamma\hat{R}_k)}} {\mbox{tr}[e^{-\beta(\hat{H}_{k}-\gamma \hat{R}_k)}]} = \frac{e^{\beta \gamma\hat{R}_k }\hat{\rho}_k}{\mbox{tr}[e^{\beta \gamma\hat{R}_k}\hat{\rho}_k]}=\hat \rho_{k} , \label{eq:extre2}
\end{equation}
where we have used the commutation relation $[\hat{R}_k,\hat{H}_k]=0$ between the Hermitian operators $\hat{R}_k$ and $\hat{H}_k$ following $[\hat{R}_k, \hat{\rho}_k] = [\hat{R}_k, e^{-\beta\hat{H}_k}] = 0$. 

Finally, combining Eq. \ref{MLE_iter} and Eq. \ref{eq:deltagt1}, we have shown that $M(\hat{\rho}_{k+1})-M(\hat{\rho}_{k}) \le 0 $ is a negative semi-definite quantity, which proves the theorem. 

We conclude that the quantum likelihood gradient in Eq. \ref{eq:iter} offers an efficient and collective optimization towards the MLE Hamiltonian, modifying all model parameters simultaneously. For each step of quantum likelihood gradient, the most costly calculation is on $\hat{\rho}_{k+1}$, or more precisely, the expectation value $\mbox{tr}[\hat{\rho}_{k+1}\hat{P}_{\lambda_i}]$ from $\hat{H}_{k+1}$. Fortunately, this is a routine calculation in quantum many-body physics and condensed matter physics with various tailored candidate algorithms under different scenarios. For example, we may resort to the FTTN, or the QMC approaches, which readily apply to much larger systems than brute-force exact diagonalization. Thus, we emphasize that MLE Hamiltonian learning works with evaluations of the expectation values of quantum states instead of the more expensive quantum states themselves in their entirety. 

Interestingly, MLE Hamiltonian learning also allows a more local stance. For a given Hamiltonian, the necessary expectation value of its Gibbs state $\mbox{tr}[\hat \rho \hat P_{\lambda_i}]$ takes the form:
\begin{eqnarray}
\mbox{tr}[\hat \rho \hat P_{\lambda_i}] &=& \mbox{tr}[\hat \rho^A_{eff} \hat P_{\lambda_i}], \nonumber \\
\hat \rho^A_{eff} = \mbox{tr}_{\bar A} [\hat \rho] &=& \frac{e^{-\beta \hat H^{A}_{eff}}}{\mbox{tr}[e^{-\beta \hat H^{A}_{eff}}]}, \label{eq:subregion}
\end{eqnarray}
where $\hat \rho^A_{eff}$ is the reduced density operator defined upon a relatively local subsystem $A$ still containing $\hat P_{ \lambda_i}$. The effective Hamiltonian $\hat{H}_{eff}^{A}=\hat{H}_A+\hat{V}_{eff}^{A}$ of the subregion $A$ contains the existing terms $\hat{H}_A$ within the subsystem and the effective interacting terms $\hat{V}_{eff}^{A}$ from the trace operation \cite{QBP2}. According to the conclusions of the quantum belief propagation theory \cite{QBP1, QBP2}, the locality of the interaction in the latter term $||\hat{V}_{eff}^{A}(l_A)||_F\propto (\beta/\beta_c)^{l_A/r}$, where $\beta$ ($\beta_c$) denotes the current (model-dependent critical) inverse temperature, $l_A$ is the distance between a specific site in the bulk of $A$ and the boundary of the subregion $A$, and $r$ is the maximum acting distance(diameter) of a single operator in the original Hamiltonian(similar to the k-local in the next section). Thus, when $\beta<\beta_c$(especially when $\beta\ll \beta_c$), $\hat{V}_{eff}^{A}$ is exponentially localized around the boundary of $A$, and the effective Hamiltonian in the bulk of $A$ remains the same as that of the original $\hat{H}_A$ of the entire system. Therefore, we may further boost the efficiency of MLE Hamiltonian learning by redirecting the expectation-value evaluations of the global quantum system to that of a series of local patches, as we will show in the next section.

In summary, given the quantum measurements of a thermal (Gibbs) state: $\{\hat{O}_{i}\}$, $N_i$, and $f_{\lambda_i}$, we can perform MLE Hamiltonian learning to obtain the MLE Hamiltonian via the following steps (Fig. \ref{fig:algorithm}): 
\begin{itemize}
    \item [1)] Initialization/Update: 
    
    For initialization, start with a random model Hamiltonian $\hat{H}_0$:
        \begin{equation}
        \hat{H}_0= \sum_{i}\mu_{i}\hat{O}_{i},
    \end{equation}
    or an identity Hamiltonian.

    For update, carry out the quantum likelihood gradient: 
        \begin{equation}
            \hat{H}_{k+1} = \hat{H}_k - \gamma \hat{R}_k,
        \end{equation}
        where $\hat{R}_k$ is defined in Eq. \ref{eq:mlerho} or Eq. \ref{eq:hiter}.
    \item [2)] Evaluate the properties $\mbox{tr}[\hat \rho_k \hat P_{\lambda_i}]$ of the quantum state:
        \begin{equation}
          \hat{\rho}_{k}=\frac{e^{-\beta\hat{H}_{k}}}{\mbox{tr}[e^{-\beta\hat{H}_{k}}]},  \label{eq:iter3}
        \end{equation}
        with suitable numerical methods.
    \item [3)] Check for convergence: loop back to step 1) to update, $k\rightarrow k+1$, if the relative entropy $M(\hat{\rho}_{k})-M_0 \ge \epsilon$ is above a given threshold $\epsilon$; otherwise, terminate the process, and the final $\hat{H}_k$ is the result for the MLE Hamiltonian. Here, $M_0$ is the theoretical minimum of the negative-log-likelihood function:
        \begin{equation}
            M_0=-\sum_{i,\lambda_i}\frac{N_i}{N_{tot}}p_{\lambda_i}\log{p_{\lambda_i}}.
        \end{equation}
\end{itemize}

In practice, $\hat R_k$ in Eq. \ref{eq:iter} is singular for small values of $\mbox{tr}[\hat \rho_k \hat P_{\lambda_i}]$ and may become numerically unstable, which requires a minimal or dynamical learning rate $\gamma$ to maintain the range of quantum likelihood gradient properly. Instead, we may employ a re-scaled version of $\hat R_k$:
\begin{equation}
        \tilde{\hat{R}}_{k} = \sum_{i,\lambda_i} \frac{N_i}{N_{tot}} f_{g}(p_{\lambda_i}/ \mbox{tr}[\hat \rho_k \hat{P}_{\lambda_i}]) \hat{P}_{\lambda_i}, \label{eq:hiter}
\end{equation}
where $f_{g}$ is a monotonic tuning-function:
\begin{equation}
    f_{g}(x)=\frac{gx}{x+g-1}, g>1,g\in\mathbb{N},
    \label{eq:fg}
\end{equation}
which maps its argument in $(0,\infty)$ to a finite range $(0,g)$. Such a re-scaled $\tilde{\hat{R}}_k$ regularizes the quantum likelihood gradient and allows a simple yet relatively larger learning rate $\gamma$ for more efficient MLE Hamiltonian learning. We also have $f_g(1)=1$, therefore $\tilde{\hat{R}}_k \rightarrow \hat{R}_k$ as we approach convergence. We will mainly employ $\tilde{\hat{R}}_{k}$ for our examples in the following sections. 

In addition to the negative-log-likelihood function $M(\hat{\rho}_k)$, we also consider the Hamiltonian distance as another criterion on the quality of Hamiltonian learning:
\begin{equation}
    \Delta{\vec{\mu}}_k = \frac{||\vec{\mu}_s-\vec{\mu}_{k}||_{2}}{||\vec{\mu}_s||_2},
    \label{eq:hamdis}
\end{equation}
where $\vec{\mu}_s$ and $\vec{\mu}_{k}$ are the (vectors of) coefficients \footnote{We typically perform MLE Hamiltonian learning and update the model Hamiltonian on the projection-operator basis; therefore, we transform the Hamiltonian back to the original, ordinary operator basis before $\Delta \vec{\mu}_k$ evaluations.} of the target Hamiltonian and the learned Hamiltonian after $k$ iterations, respectively. However, we do not recommend $\Delta{\vec{\mu}}_k$($\Delta \mu$ for short) as convergence criteria as $\vec{\mu}_s$ is generally unknown aside from benchmark scenarios \cite{Zoller}.

\section{Example models and results}

In this section, we demonstrate the performance of the MLE Hamiltonian learning protocol. For better numerical simulations, we consider the $k$-local Hamiltonians, with operators acting non-trivially on no more than $k$ contiguous sites in each direction. For example, for a 1-dimensional spin-$\frac{1}{2}$ system, a $k$-local operator for $k=2$ takes the form $\hat{S}_{i}^{\alpha}\hat{S}_{i+1}^{\beta}$ or $\hat{S}_{i}^{\alpha}$, $\alpha, \beta\in\{x,y,z\}$, where $\hat{S}_{i}^{\alpha}$ denotes the spin operator. In particular, we focus on general 1D quantum spin chains with $k=2$, taking the following form:
\begin{equation}
    \hat{H}_s = \sum_{i,\alpha,\beta}^{L-1}J_{i}^{\alpha\beta}\hat{S}_{i}^{\alpha}\hat{S}_{i+1}^{\beta}+\sum_{i,\alpha}^{L}h_{i}^{\alpha}\hat{S}_{i}^{\alpha},
    \label{XZ}
\end{equation}
where $\hat{S}_{i}^{\alpha}$ denotes the spin operator on site $i$, $\alpha,\beta,\in\{x,y,z\}$. There are $12L-9$ 2-local operators under the open boundary condition, where $L$ is the system size. We generate the model parameters $\vec{\mu}_s = \{ J_{i}^{\alpha\beta}, h_{i}^{\alpha} \}$ randomly following a uniform distribution in $[-1,1]$. This Hamiltonian $\hat{H}_s$, specifically the model parameters $\vec{\mu}_s$, will be our target for MLE Hamiltonian learning. As the protocol's inputs, we simulate quantum measurements of all 2-local operators $\{\hat{O}_i\}$ on the Gibbs states of $\hat{H}_s$ numerically via exact diagonalization on small systems and FTTN for large systems. For the latter, we use a tensor network ansatz called the ``ancilla" method \cite{FTTN}, where we purify a Gibbs state with some auxiliary qubits $\hat\rho_s = \mbox{tr}_{aux}|\psi_s\rangle\langle \psi_s|$, and obtain $\ket{\psi_s}=e^{-\beta\hat{H}_s/2}\ket{\psi_0}$ from a maximally-entangled state $\ket{\psi_0}$ via imaginary time evolution. In addition, given a large number $n$ of Trotter steps, the imaginary time evolution operator $e^{-\beta\hat{H}_s/2}$ is decomposed into Trotter gates' product as $(\Pi_{i}e^{-\mu_i\hat{O}_{i} \beta/2n})^{n}+O(\beta^2/n)$. Here, we set the Trotter step $\delta t=\beta/n \in [0.01,0.1]$, for which the Trotter errors of order $O(\beta^2/n)$ show little impact on our protocol's accuracy. Without loss of generality, we employ the integrated FTTN algorithm in the ITensor numerical toolkit \cite{ITensor}, and set the number of measures $N_i=N$ for all operators in our examples for simplicity.

\begin{figure}
    \centering
    \includegraphics[width = 1\linewidth]{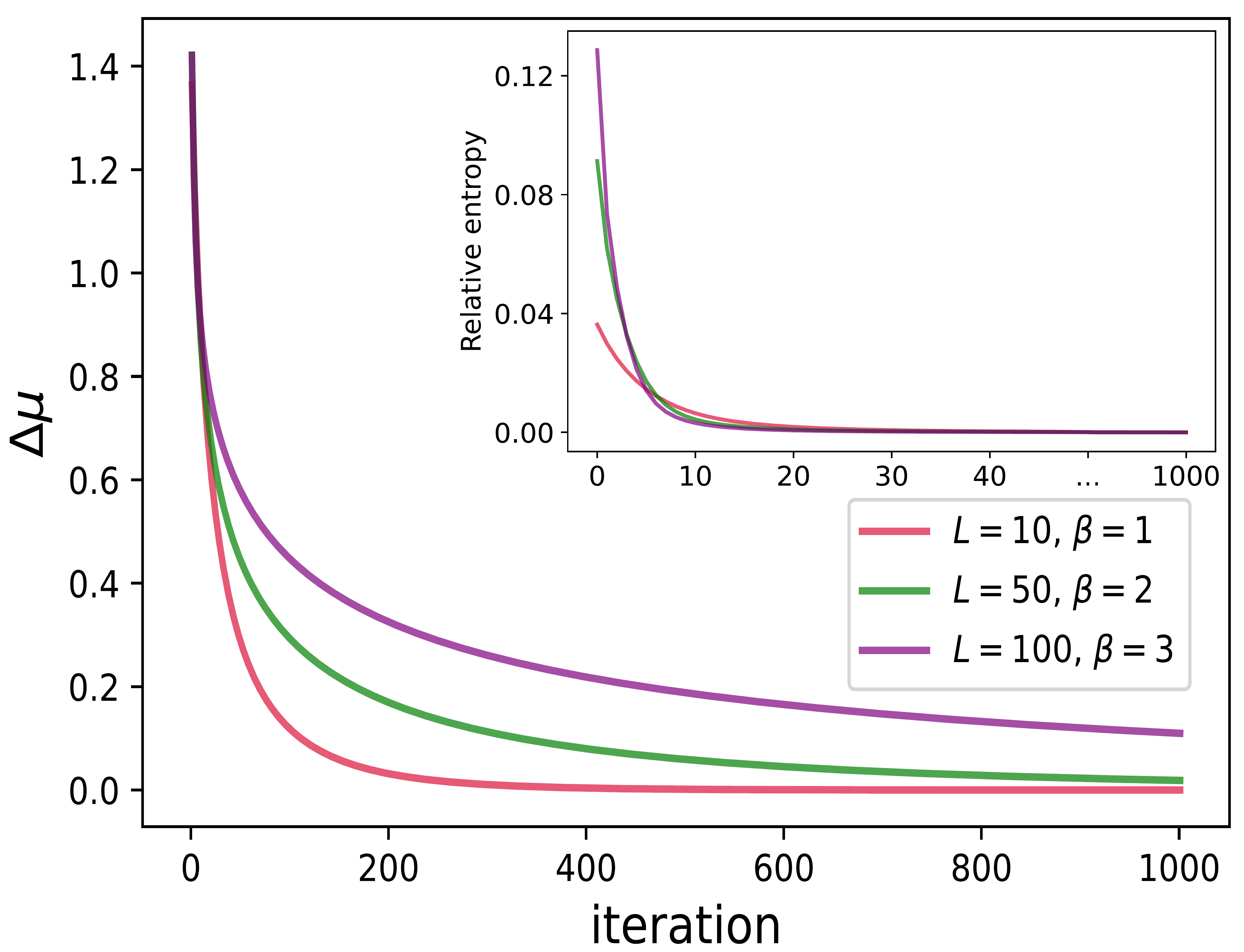}
    \caption{Both the Hamiltonian distance $\Delta\mu$ defined in Eq. \ref{eq:hamdis} and the negative-log-likelihood function $M(\hat{\rho}_{k+1})$ (or relative entropy $M(\hat{\rho}_{k+1})-M_0$) show successful convergence of the iterations in MLE Hamiltonian learning, albeit a variety of system sizes and temperature ranges. We simulate the target Hamiltonian and the iteration process by FTTN with Trotter step $\delta t=0.1$. Each curve is averaged on 10 trials of random $\hat{H}_0$ initializations. We set the learning rate $\gamma=0.1$. The maximum number of iterations here is 1000.}
    \label{Gibbs_differ}
\end{figure}

As we demonstrate in Fig. \ref{Gibbs_differ}, MLE Hamiltonian learning obtains the target Hamiltonians with high accuracy and efficiency under various settings of system sizes and inverse temperatures $\beta$. Besides, instead of the original quantum likelihood gradient in Eq. \ref{eq:mlerho}, we may obtain a faster convergence with the re-scaled $\tilde{\hat{R}}_k$ in Eq. \ref{eq:hiter} and a larger learning rate, as we discuss in Appendix B. In the following numerical examples, we use the re-scaled quantum likelihood gradient $\tilde{\hat{R}}_k$ and set $g=2$ for the tuning function in Eq. \ref{eq:fg}. Within the given iterations, not only have we achieved results (Hamiltonian distance $\Delta\mu\sim O(10^{-12})$ and relative entropy $M(\hat{\rho}_k)-M_0\sim O(10^{-16})$) comparable to, if not exceeding, previous methods \cite{Corr1} for $L=10$ systems and $\beta=1$ straightforwardly, but we have also achieved satisfactory consistency ($\Delta\mu\sim O(10^{-2})$ and $M(\hat{\rho}_k)-M_0\sim O(10^{-9})$) for large systems $L=100$ and low temperatures $\beta=3$ that were previously inaccessible.

\begin{figure}
    \centering
    \includegraphics[width = 1\linewidth]{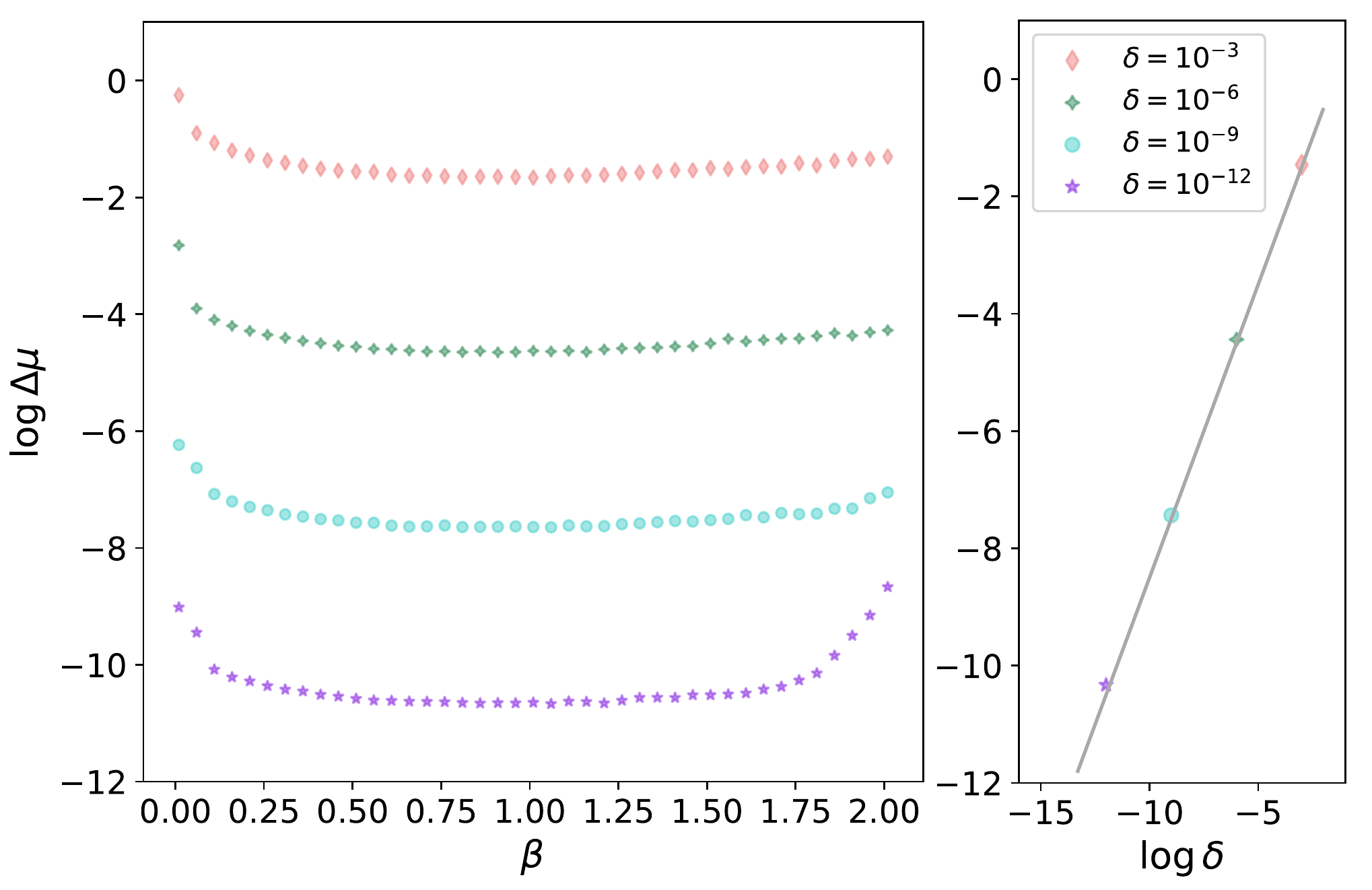}
    \caption{The performance of MLE Hamiltonian learning maintains relatively well against noises and, especially, broader temperature ranges. Left: the Hamiltonian distance versus the inverse temperature $\beta$ shows a broader applicable temperature range. Each data point contains 10 trials. Right: the performance (left figure's data averaged over temperature) versus the noise strength $\delta$ shows the impact of noises and the protocol's relative robustness against them. The slope of the straight line is $\sim 1$, indicating a linear relationship between $\Delta\mu$ and $\delta$. Note the log scale $\log(\Delta\mu)$ for the vertical axis. We set $L=10$ for the system size, and learning rate $\gamma=1$.}
    \label{Gibbs_temp}
\end{figure}

MLE Hamiltonian learning is also relatively robust against temperature and noises, two key factors impacting accuracy in Hamiltonian learning. For illustration, we include random errors $\delta \langle \hat O_i \rangle$ following Gaussian distribution with zero mean and standard deviation $\delta$ to all quantum measurements: $\langle \hat O_i \rangle \rightarrow \langle \hat O_i \rangle + \delta \langle \hat O_i \rangle$. We note that such $\delta$ may also depict the quantum fluctuations \cite{Corr1, Corr2} from a finite number of measurements $\delta \propto N_i^{-1/2}$. We also focus on smaller systems with $L=10$ and employ exact diagonalization to avoid confusion from potential Trotter error of the FTTN ansatz\cite{FTTN}. We summarize the results in Fig. \ref{Gibbs_temp}. 

Most previous algorithms on Hamiltonian learning have a rather specific applicable temperature range. For example, the high-temperature expansion of $e^{-\beta\hat{H}}$ only works in the $\beta\ll 1$ limit \cite{ogl, PhysRevA.92.052322}. Besides, gradient descent on the log partition function, despite a convex optimization, performs well in a narrow temperature range \cite{RN571}. The gradient of this algorithm is proportional to the inverse temperature, so the algorithm's convergence slows at high temperatures. Also, the gradient descent algorithm cannot extend to the $\beta\rightarrow\infty$ limit - the ground state, while our protocol is directly applicable to the ground states of quantum systems, as we will generalize and justify later.

MLE Hamiltonian learning is also more robust to noises, with an accuracy of Hamiltonian distance $\Delta\mu \sim O(10^{-11})$ across a broad temperature range at noise strength $\delta \sim O(10^{-12})$. Such noise level is hard to realize in practice; nevertheless, it is necessary to safeguard the correlation matrix method \cite{Qi2019, Corr1, sbhl}. Even so, due to the uncontrollable spectral gap, the correlation matrix method is susceptible to high temperature, and its accuracy drastically decreases to $\Delta\mu \sim O(10^{-3})$ at $\beta=0.01$. In comparison, MLE Hamiltonian learning is more versatile, with an approximately linear dependence between its accuracy $\Delta\mu$ and the noise strength $\delta$ across a broad range of temperatures and noise strengths, saturating the previous bound \cite{RN571}; see the right panel of Fig. \ref{Gibbs_temp}. We also provide more detailed comparisons between the algorithms in Appendix C.

\begin{figure}
    \centering
    \includegraphics[width = 0.98\linewidth]{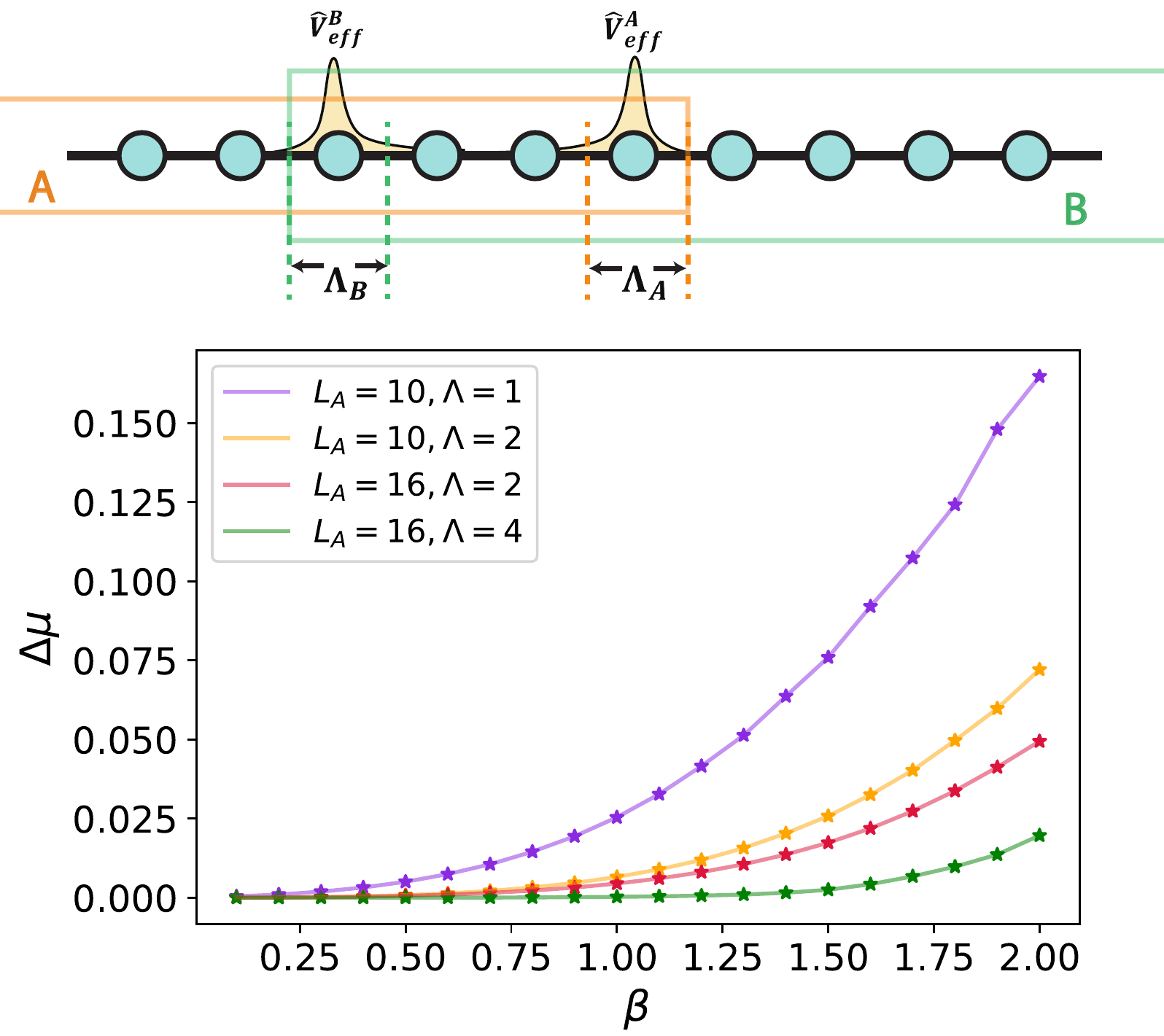}
    \caption{Upper: MLE Hamiltonian learning's need for evaluations of local observables can be satisfied among local patches for thermal states at sufficiently high temperatures, where the effective potential $V^A_{eff}$ ($V^{B}_{eff}$) is weak and localized on the boundaries of subregion $A$ ($B$) within a cut-off range $\Lambda_A$ ($\Lambda_B$). Consequently, for $\hat{P}_{\lambda_i}$ defined sufficiently deep inside $A$ ($B$), we can estimate its expectation value via $\mbox{tr}[\hat{\rho}^{A}\hat{P}_{\lambda_i}]$ ($\mbox{tr}[\hat{\rho}^{B}\hat{P}_{\lambda_i}]$). Lower: the Hamiltonian distance $\Delta\mu$ of the results after MLE Hamiltonian learning indicates better validity of the local-patch approximation at higher temperatures. The total system size is $L=100$, while the local patches are of sizes $L_A=10, 16$ with cut-offs $\Lambda=1, 2, 4$, respectively. Each data point contains 10 trials. We set $\delta t=0.1$ for the Trotter step in FTTN ansatz and learning rate $\gamma=1$. }
    \label{subregion}
\end{figure}

Despite efficient quantum likelihood gradient and applicable quantum many-body ansatz, the computational cost of MLE Hamiltonian learning still increases rapidly with the system size $L$. Fortunately, as stated above in Eq. \ref{eq:subregion}, we may resort to calculations on local patches, especially for low dimensions and high temperatures due to their quasi-Markov property. In particular, when $\beta<\beta_c$ ($T>T_c$), the difference between the cutoff Hamiltonian $\hat{H}_A$ and the effective Hamiltonian $\hat{H}_{eff}^{A}$ in a local subregion $A$, $V_{eff}^{A}$, should be weak, short-ranged, and localized at $A$'s boundary \cite{QBP1, QBP2}; therefore, for those operators $\hat{P}_{\lambda_i}$ adequately deep inside $A$, we can use $\hat \rho^A$, the Gibbs state defined by $\hat{H}_A$, to estimate the corresponding $\mbox{tr}[\hat{\rho}\hat{P}_{\lambda_i}]$; see illustration in Fig. \ref{subregion} upper panel. 

For example, we apply MLE Hamiltonian learning on $L=100$ systems, where we iteratively calculate the necessary expectation values on different local patches of size $L_A=10, 16$. We also choose different cut-offs $\Lambda$, and evaluate $\mbox{tr}[\hat{\rho}^{A}\hat{P}_{\lambda_i}]$ for those operators at least $\Lambda$ away from the boundaries and sufficiently deep inside the subregion $A$, so that the effective potential $V^{A}_{eff}$ may become negligible. We also employ a sufficient number of local patches to guarantee full coverage of necessary observables - operators outside $A$ or in $\Lambda_A$ are obtainable from another local patch $B$, as shown in the upper panel of Fig. \ref{subregion}, and so on so forth. Both the $L=100$ target system and the local patches for MLE Hamiltonian learning are simulated via FTTN. We have no problem achieving convergence, and the resulting Hamiltonians' accuracy, the Hamiltonian distance $\Delta\mu$ versus the inverse temperature $\beta$, is summarized in the lower panel of Fig. \ref{subregion}. Indeed, the local-patch approximation is more reliable at higher temperatures, as well as with larger subsystems and cutoffs, albeit with rising costs. We also note that we can achieve much larger systems with the local patches than $L=100$ we have demonstrated.

\section{MLE Hamiltonian learning for pure eigenstates}

In addition to the Gibbs states, MLE Hamiltonian learning also applies to measurements of certain eigenstates of target quantum systems:

1. The ground states are essentially the $\beta \rightarrow \infty$ limit of the Gibbs states. However, due to the order-of-limit issue, the $\gamma \rightarrow 0$ requirement of the theorem on Gibbs states forbids a direct extension to the ground states. In the Appendix D, we offer rigorous proof of the effectiveness of quantum likelihood gradient based on ground-state measurements, along with several nontrivial MLE Hamiltonian learning examples on quantum critical and topological ground states. We note that Ref. \onlinecite{wjb} offers preliminary studies on pure-state quantum state tomography, inspiring this work. 

2. A highly-excited eigenstate of a (non-integrable) quantum chaotic system $\hat{H}_s$ is believed to obey the eigenstate thermalization hypothesis (ETH), that its density operator $\hat{\rho}_{s}=\ket{\psi_s}\bra{\psi_s}$ behaves locally indistinguishable from a Gibbs state $\hat \rho_{s, A}$ in thermal equilibrium \cite{Tarun}:
\begin{equation}
    \hat{\rho}_{s,A}=\mbox{tr}_{\bar{A}}[\hat{\rho}_s]\approx \frac{e^{-\beta_s\hat{H}_A}}{\mbox{tr}[e^{-\beta_s\hat{H}_A}]},
    \label{eq:eth}
\end{equation}
where $\beta_s$ is an effective temperature determined by the energy expectation value $\braket{\psi_s|\hat{H}_s|\psi_s}=\frac{\mbox{tr}[e^{-\beta_s\hat{H}_s}\hat{H}_s]}{\mbox{tr}[e^{-\beta_s\hat{H}_s}]}$. As MLE Hamiltonian learning only engages local operators, its applicability directly generalizes to such eigenstates $\ket{\psi_s}$ following ETH. 

3. In general, ETH applies to eigenstates in the center of the spectrum of quantum chaotic systems, while low-lying eigenstates are too close to the ground state to exhibit ETH \cite{PhysRevE.90.052105}. However, in the rest of the section, we demonstrate numerically that MLE Hamiltonian learning still works well for low-lying eigenstates. 

We consider the 1D longitudinal-transverse-field Ising model \cite{Tarun, PhysRevE.90.052105} as our target quantum system: 
\begin{equation}
    \hat{H}_s=J\sum_{j}^{L-1}\hat{\sigma}_{j}^{z}\hat{\sigma}_{j+1}^{z}+g_{z}\sum_{j}^{L}\hat{\sigma}_{j}^{z}+g_{x}\sum_{j}^{L}\hat{\sigma}_{j}^{x},
    \label{ETH_ham}
\end{equation}
where the system size is $L=80$. We set $J=1$, $g_{x}=0.9045$, and $g_{z}=0.8090$. The quantum system is strongly non-integrable under such settings. Previous studies mainly focused on eigenstates in the middle of the energy spectrum. In contrast, we pick the first excited state - a typical low-lying eigenstate considered asymptotically integrable and ETH-violating \cite{PhysRevE.90.052105} - for quantum measurements (via DMRG) and then MLE Hamiltonian learning for its candidate Hamiltonian (via FTTN). 

\begin{figure}
    \centering
    \includegraphics[width = 0.98\linewidth]{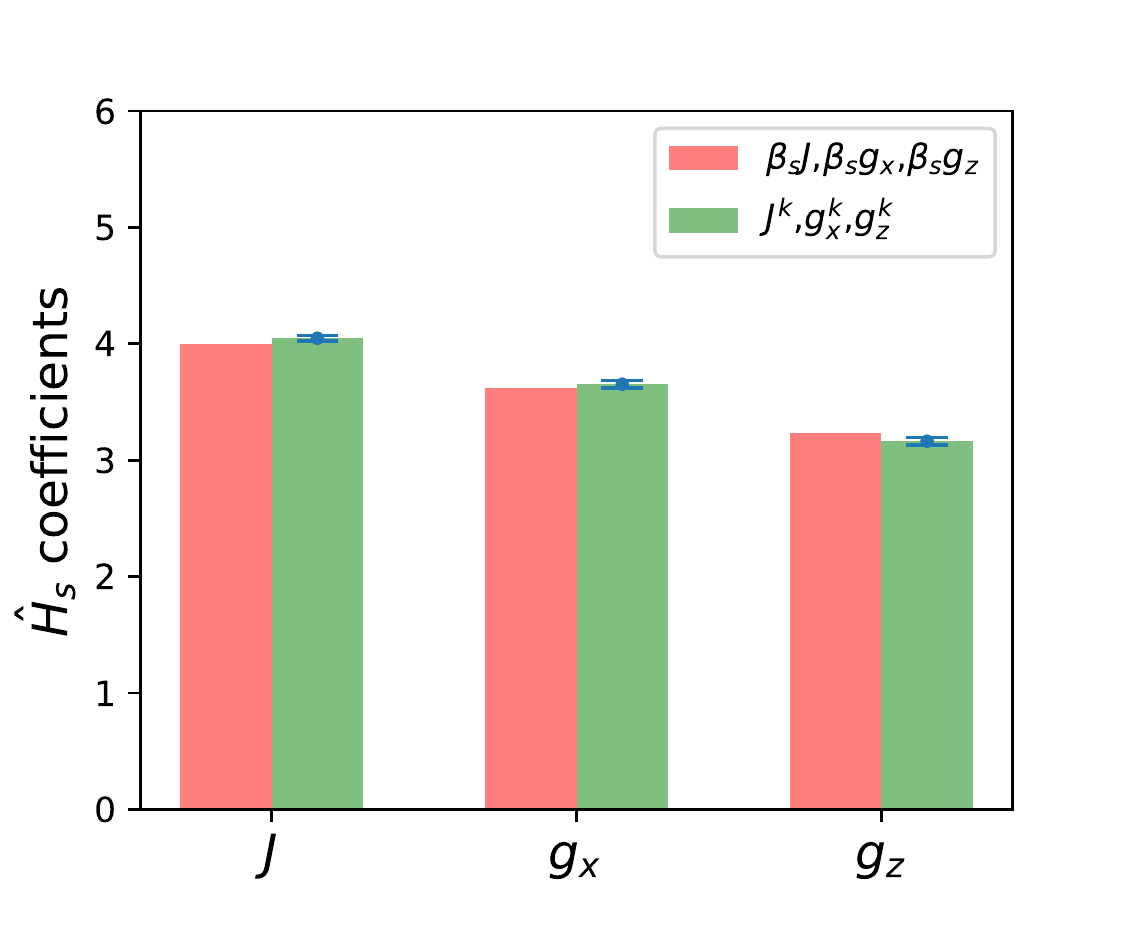}
    \caption{The coefficients obtained via MLE Hamiltonian learning (green columns) compare well with those of the target Hamiltonian $\hat{H}_s$ even though the quantum measurements are based upon a low-lying (first) excited state. The red columns denote the coefficients of $\hat{H}_s$ in Eq. \ref{ETH_ham} multiply the effective (inverse) temperature $\beta_s=4$. The error bars demonstrate the variances over the lattices and trials.  We set the system size $L=80$, learning rate $\gamma=0.1$, and the Trotter step $\delta t=0.1$.}
    \label{ETH}
\end{figure}

We summarize the results in Fig. \ref{ETH}. Further, the model Hamiltonian we established is approximately equivalent to the target quantum Hamiltonian at an (inverse) temperature $\beta_s\approx 4$ \cite{Tarun}, which we have absorbed into the unit of our $\hat{H}_k$. Therefore, we have accurately established the model Hamiltonian and derived the effective temperature consistent with previous results \cite{Tarun} for a low-lying excited eigenstate not necessarily following ETH. The physical reason for quantum likelihood gradient applicability in such states is an interesting problem that deserves further studies.

\section{Discussions}

We have proposed a novel MLE Hamiltonian learning protocol to achieve the model Hamiltonian of the target quantum system based on quantum measurements of its Gibbs states. The protocol updates the model Hamiltonian iteratively with respect to the negative-log-likelihood function from the measurement data. We have theoretically proved the efficiency and convergence of the corresponding quantum likelihood gradient and demonstrated it numerically on multiple non-trivial examples, which show more accuracy, better robustness against noises, and less temperature dependence. Indeed, the accuracy is almost linear to the imposed noise amplitude, thus inverse proportional to the square root of the number of samples, the asymptotic upper bound\cite{RN571}. Further, MLE Hamiltonian learning directly rests on the Hamiltonians and their physical properties instead of direct and costly access to the quantum many-body states. Consequently, we can resort to various quantum many-body ansatzes in our systematic quantum toolbox and even local-patch approximation when the situation allows. These advantages allow applications to larger systems and lower temperatures with better accuracy than previous approaches. On the other hand, while our protocol is generally applicable for learning any Hamiltonian, its advantages are most apparent for local Hamiltonians, where various quantum many-body ansatzes and local-patch approximation shine. Despite such limitations, we note that the physical systems are characterized by local Hamiltonians in a significant proportion of scenarios. 

In addition to the Gibbs states, we have generalized the applicability of MLE Hamiltonian learning to eigenstates of the target quantum states, including ground states, ETH states, and even selected cases of low-lying excited states. We have also provided theoretical proof of quantum likelihood gradient rigor and convergence in the Appendix D, along with several other numerical examples. 

Our strategy may apply to the entanglement Hamiltonians and the tomography of the quantum states under the maximum-likelihood-maximum-entropy assumption \cite{PhysRevLett.107.020404}. Besides, our algorithm may also provide insights into the quantum Boltzmann machine \cite{QBM} - a quantum version of the classical Boltzmann machine with degrees of freedom that obey the distribution of a target quantum Gibbs state. Instead of brute-force calculations of the loss function derivatives with respect to the model parameters or approximations with the gradients' upper bounds, our protocol provides an efficient optimization that updates the model parameters collectively.

\emph{Acknowledgement:}- We thank insightful discussions with Jia-Bao Wang. We acknowledge support from the National Key R\&D Program of China (No.2021YFA1401900) and the National Science Foundation of China (No.12174008 \& No.92270102). The calculations of this work are supported by HPC facilities at Peking University. 

\bibliography{refs.bib}

\newpage

\appendix

\section{Maximum condition for MLE}

In this appendix, we review the derivation of the maximum condition \cite{PhysRevA.75.042108} in Eq. \ref{eq:mlerho} in the main text. 

A general quantum state takes the form of a density operator:
\begin{equation}
    \hat{\rho}=\sum_{j}p_j\ket{\psi_j}\bra{\psi_j},
    \label{general}
\end{equation}
where $p_j\ge 0$, $\sum_j p_j=1$, and $\ket{\psi_j}$ is a set of orthonormal basis. The search for the quantum state that maximizes the likelihood function:
\begin{equation}
    \mathcal{L}(\hat{\rho}) = \prod_{i, \lambda_i}\{\mbox{tr}[\hat{\rho}\hat{P}_{\lambda_i}]^{\frac{f_{\lambda_i}}{N_{tot}}}\}^{N_{tot}},
\end{equation}
can be converted to the optimization problem:
\begin{equation}
    \begin{split}
        &\min\limits_{\hat{\rho}\in\mathcal{D}}\quad M(\hat{\rho})=-\frac{1}{N_{tot}}\log[\mathcal{L}(\hat{\rho})]\\
        &{\rm subject\enspace to}\quad \hat{\rho}\succeq 0, \mbox{tr}[\hat{\rho}]=1.\\
        \label{eq:optprob}
    \end{split}
\end{equation}
It is hard to solve this semi-definite programming problem directly and numerically. Instead, forgoing the non-negative definiteness, we adopt the Lagrangian multiplier method:
\begin{equation}
    \frac{\partial}{\partial{\bra{\psi_j}}}\{M(\hat{\rho})+\lambda \mbox{tr}[\hat{\rho}]\}=0,
\end{equation}
where $\lambda$ is a Lagrangian multiplier. Given Eq. \ref{general}, we obtain the following solution:
\begin{equation}
\begin{split}
        &\hat{R}\ket{\psi_j}=\ket{\psi_j}\\
        &\hat{R} = \sum_{i,\lambda_i}\frac{f_{\lambda_i}}{N_{tot}}  \frac{\hat{P}_{\lambda_i}}{\mbox{tr} [\hat \rho \hat P_{\lambda_i}]},\\
\end{split}
\label{extre_appendix}
\end{equation}
and $\lambda=1$. Combining Eq. \ref{general} and Eq. \ref{extre_appendix}, we obtain the maximum condition:
\begin{equation}
    \hat{R}\hat{\rho}=\hat{\rho}.
    \label{eq:exteme}
\end{equation}
We note that Eq. \ref{extre_appendix} does not guarantee the positive semi-definiteness of the density operator. Instead, one may search within the density-operator space (the space of positive semi-definite matrix with unit trace) to locate the MLE quantum state fulfilling Eq. \ref{extre_appendix} or Eq. \ref{eq:exteme}. For the Hamiltonian learning task in this work, the search space is naturally the space of Gibbs states (under selected quantum many-body ansatz).

\section{MLE Hamiltonian learning with rescaling function}

\begin{figure}
    \centering
    \includegraphics[width = 1\linewidth]{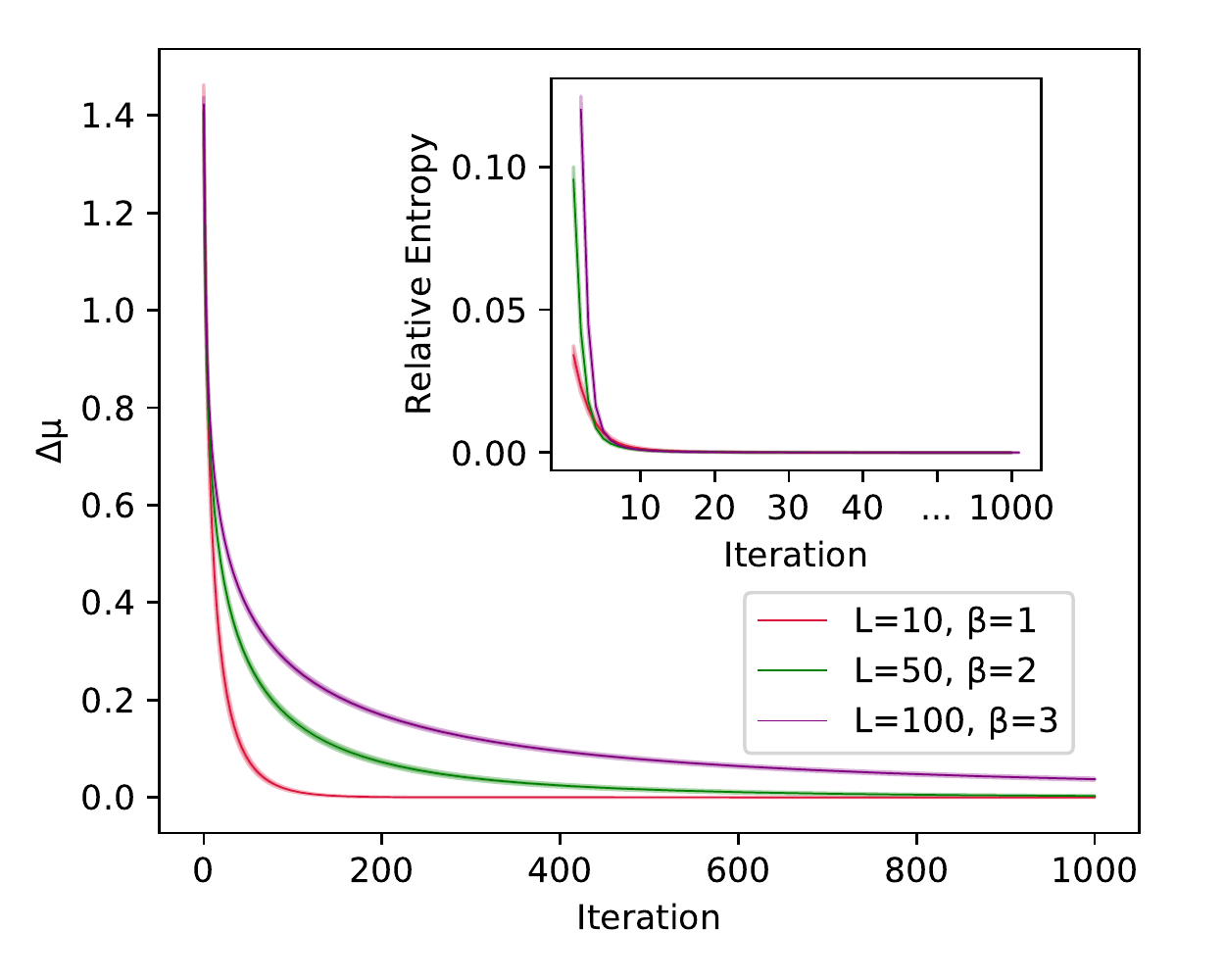}
    \caption{Both the Hamiltonian distance $\Delta\mu$ defined in Eq. \ref{eq:hamdis} and the negative-log-likelihood function $M(\hat{\rho}_{k+1})$ (or relative entropy $M(\hat{\rho}_{k+1})-M_0$) show successful convergence of the iterations in MLE Hamiltonian learning, albeit a variety of system sizes and temperature ranges. We simulate the target Hamiltonian and the iteration process by FTTN with Trotter step $\delta t=0.1$. Each curve is averaged on 10 trials of random $\hat{H}_0$ initializations. The maximum number of iterations here is 1000. In comparison with Fig. 2 in the main text, we use the re-scaled $\tilde{\hat{R}}_k$ in Eq. 16 in the main text with $g=2$, which allows us to employ a larger learning rate $\gamma=1$ and obtain a faster convergence. }
    \label{fig:rescaled_FIG2}
\end{figure}

In this appendix, we compare the MLE Hamiltonian learning with the quantum likelihood gradient $\hat{R}_k$ and the re-scaled counterpart $\tilde{\hat{R}}_k$. As we state in the main text, $\tilde{\hat{R}}_k$ regularizes the gradient, allowing us to employ a larger learning rate $\gamma=1$, which leads to a faster convergence (Fig. \ref{fig:rescaled_FIG2}) and a higher accuracy (Tab. \ref{table}) given identical number of iterations. 

\begin{table}
\renewcommand\arraystretch{1.6}
\resizebox{80mm}{!}{
\begin{tabular}{|c|c|c|c|r|l|} \hline 
 & $L=10$,$\beta=1$ & $L=50$,$\beta=2$ & $L=100$,$\beta=3$  \\ \hline
$\hat{R}_k$ &$O(10^{-6})$ &	$O(10^{-2})$ &	$O(10^{-1})$ \\ \hline
$\tilde{\hat{R}}_k$ & $O(10^{-12})$ & $O(10^{-3})$ & $O(10^{-2})$ \\ \hline
\end{tabular}}
\caption{The algorithm's accuracy (Hamiltonian distance $\Delta\mu$) further improves with a re-scaled quantum likelihood gradient $\tilde{\hat{R}}_k$ under various system sizes $L$ and (inverse) temperatures $\beta$. }
\label{table}
\end{table}

\section{Comparisons between Hamiltonian learning algorithms}

In this appendix, we compare different Hamiltonian learning algorithms, including the correlation matrix (CM) method \cite{Qi2019, Corr1}, the gradient descent (GD) method \cite{RN571}, and the MLE Hamiltonian learning (MLEHL) algorithm, by looking into some of their numerical results and performances. We consider general 2-local Hamiltonians in Eq. \ref{XZ} in the main text for demonstration and measurements $\{\hat{O}_i\}$ over all the 2-local operators (instead of all 4-local operators as in Ref. \cite{Corr1}).

\begin{figure}
    \centering
    \includegraphics[width = 1\linewidth]{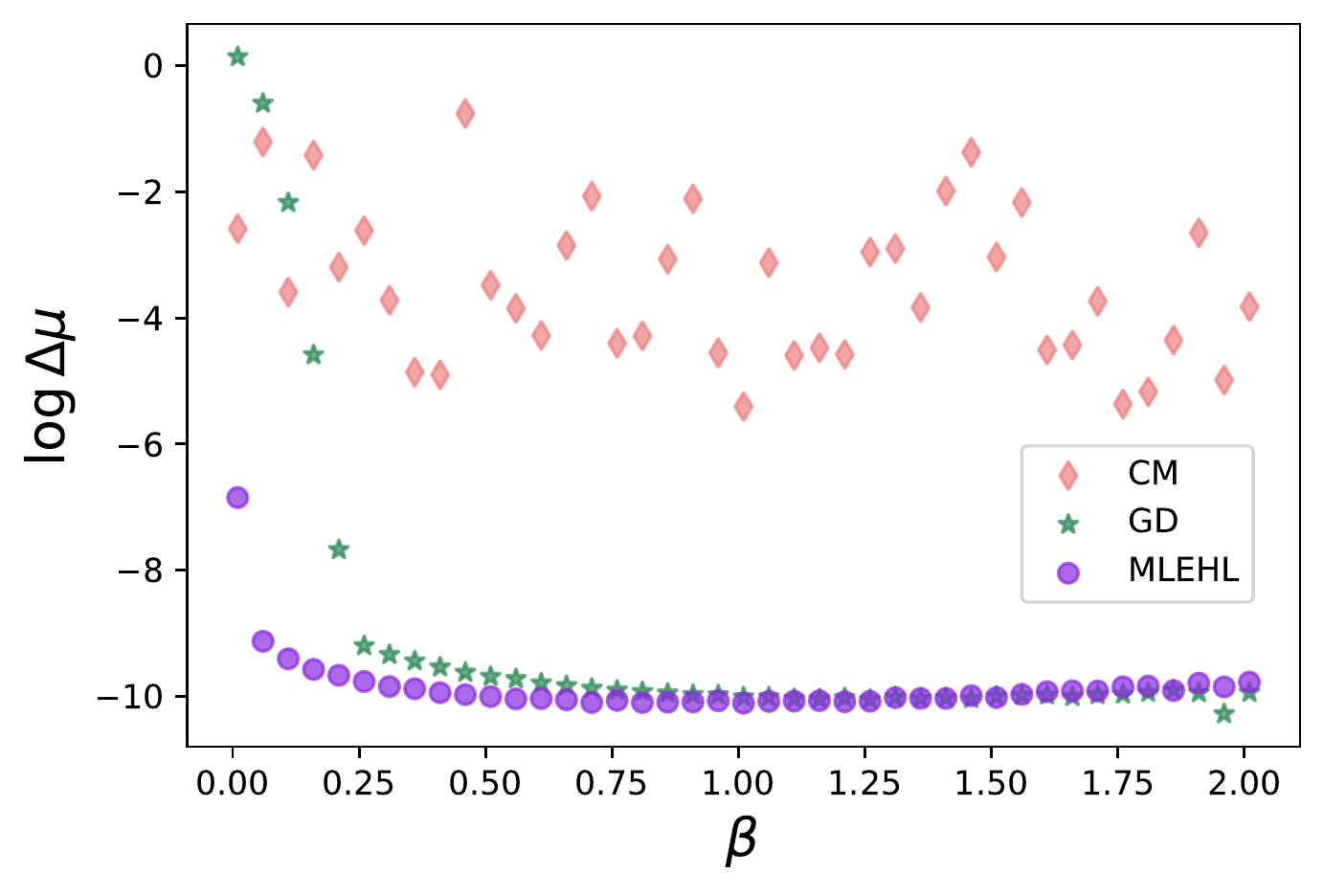}
    \caption{The performances (logarithm of Hamiltonian distances) of different algorithms versus the inverse temperature $\beta$ show the advantages of the MLEHL algorithm. Each data point contains 10 trials of random Hamiltonians. We also include noises following a normal distribution with zero means and $O(10^{-12})$ standard deviation. For both GD and MLEHL algorithms, we employ a learning rate $\gamma = 1$ and a maximum number of iterations of 7000. The system size is $L=7$.}
    \label{fig:compare}
\end{figure}

We summarize the results in Fig. \ref{fig:compare}: the accuracy of CM is unstable and highly sensitive to temperature; while GD performs similarly to the proposed MLEHL algorithm at low temperatures, its descending gradient becomes too small at high temperatures to allow a satisfactory convergence within the given maximum iterations. 

\begin{figure}
    \centering
    \includegraphics[width = 1\linewidth]{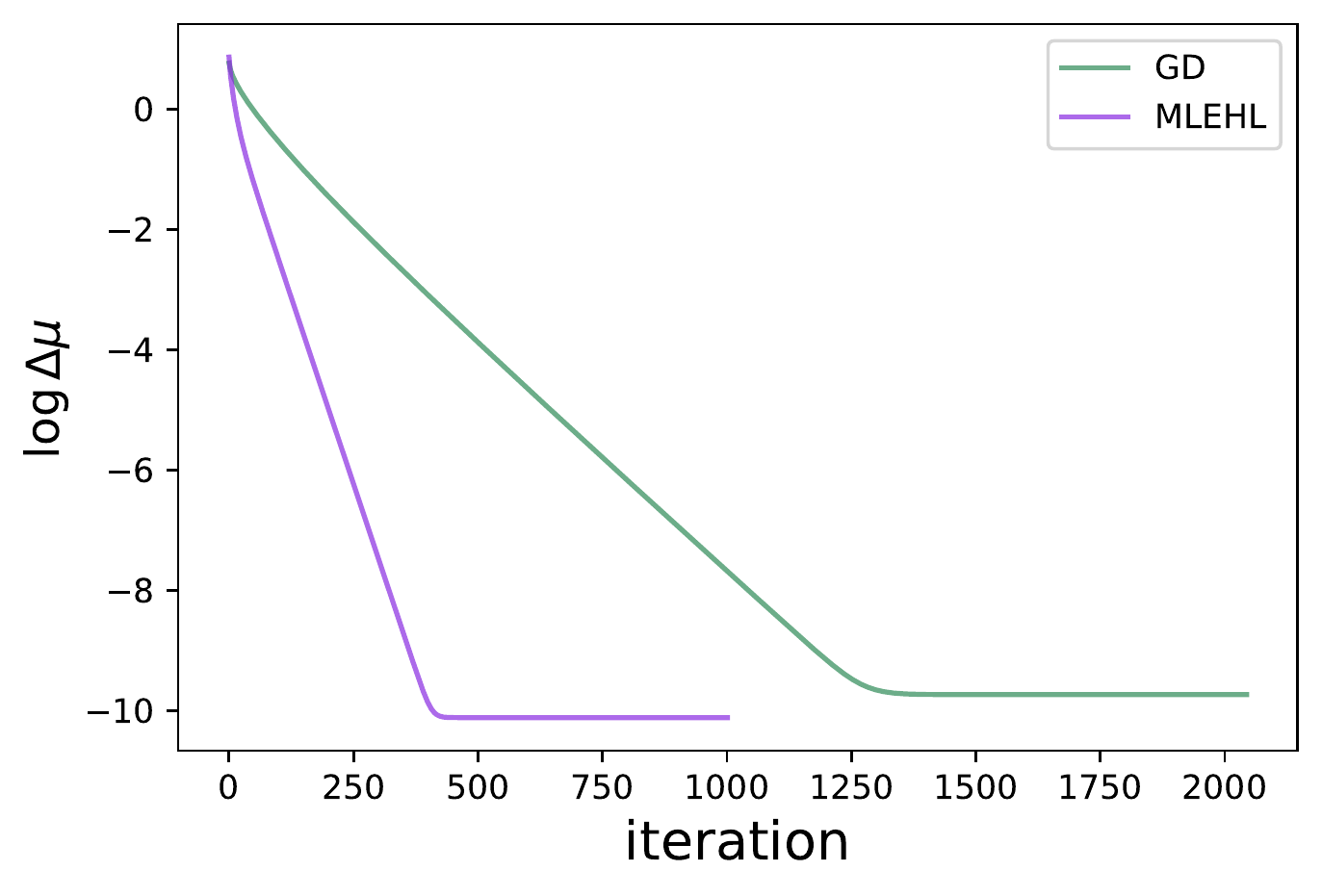}
    \caption{The logarithm of the Hamiltonian distances versus the number of iterations shows a faster convergence under the proposed MLEHL algorithm. We also include noises following a normal distribution with zero means and $O(10^{-12})$ standard deviation. We employ a learning rate of $\gamma = 1$ for both GD and MLEHL. We set the system size $L=7$ and the (inverse) temperature $\beta=1$.}
    \label{fig:compare_GD_MLEHL}
\end{figure}

We also compare the convergence rates of the MLEHL and GD algorithms with the same learning rate. As in Fig. \ref{fig:compare_GD_MLEHL}, the MLEHL algorithm exhibits a faster convergence and a smaller computational cost, which is similar under both algorithms for each iteration.  

\section{Hamiltonian learning from ground state}

In this appendix, we prove the effectiveness of the quantum likelihood gradient based on measurements of the target quantum system's ground state and provide several nontrivial numerical examples, including 1D quantum critical states and 2D topological states.

\subsection{Proof for ground-state-based quantum likelihood gradient}

Given a sufficient number $N_i$ measurements of the operator $\hat{O}_i$ on the non-degenerate ground state $\ket{\psi_s}$ of a target system $\hat{H}_s$, we obtain a number of outcomes as the $\lambda_i^{th}$ eigenvalue of $\hat{O}_i$ as:
\begin{equation}
f_{\lambda_i}=p_{\lambda_i}N_i\approx\bra{\psi_s}\hat{P}_{\lambda_i}\ket{\psi_s}N_i,
\end{equation}
where $p_{\lambda_i}=f_{\lambda_i}/N_i$, and $\hat{P}_{\lambda_i}$ is the projection operator of the eigenvalue $o_{\lambda_i}$.

Our MLE Hamiltonian learning follows the iterations:
\begin{equation}
    \begin{split}
        &\hat{H}_{k+1}=\hat{H}_k-\gamma\hat{R}_k,\\
        &\hat{R}_k = \sum_{i,\lambda_i}\frac{f_{\lambda_i}}{N_{tot}}\frac{\hat{P}_{\lambda_i}}{\bra{\psi_{k}^{gs}}\hat{P}_{\lambda_i}\ket{\psi_{k}^{gs}}},\\
    \end{split}
    \label{iter_gs}
\end{equation}
where $\ket{\psi_k^{gs}}$ is the non-degenerate ground state of $\hat{H}_k$. 

\textbf{Theorem:} For $\gamma\ll1,\gamma>0$, the quantum likelihood gradient in Eq. \ref{iter_gs} yields a negative semi-definite contribution to the negative-log-likelihood function $M(\ket{\psi_{k+1}^{gs}})=-\frac{1}{N_{tot}}\log\mathcal{L}(\ket{\psi_{k+1}^{gs}})$ following Eq. \ref{LHF} in the main text.

\textbf{Proof:} At the linear order in $\gamma$, we may treat the addition of $-\gamma\hat{R}_k$ to $\hat{H}_k$ at the $k^{th}$ iteration as a perturbation:
\begin{equation}
        \ket{\psi_{k+1}^{gs}}=\ket{\psi_{k}^{gs}}-\gamma\hat{G}_{k}\hat{R}_{k}\ket{\psi_{k}^{gs}}+O(\gamma^2),\label{eq:itergs_pertb}
\end{equation}
where $\hat{G}_k$ is the Green's function in the $k_{th}$ iteration:
\begin{equation}        
        \hat{G}_k =\hat{Q}_k \frac{1}{E_{k}^{gs}-\hat{H}_k}\hat{Q}_k,
\end{equation}
where $\hat{Q}_k=I-\ket{\psi_{k}^{gs}}\bra{\psi_{k}^{gs}}$ is the projection operator orthogonal to the ground space $\ket{\psi_{k}^{gs}}\bra{\psi_{k}^{gs}}$, and $E_{k}^{gs}$ is the ground state energy. Keeping terms upto the linear order of $\gamma$ in the log expansion of the negative-log-likelihood function, we have:
\begin{equation}
    \begin{split}
        M(\ket{\psi_{k+1}^{gs}})&=-\frac{1}{N_{tot}}\log\mathcal{L}(\ket{\psi_{k+1}^{gs}})\\
        &=-\sum_{i,\lambda_i}\frac{f_{\lambda_i}}{N_{tot}}\log\bra{\psi_{k+1}^{gs}}\hat{P}_{\lambda_i}\ket{\psi_{k+1}^{gs}},\\
        &\approx M(\ket{\psi_k^{gs}})+2\gamma\Delta_k.
    \end{split}
    \label{nlf_gs}
\end{equation}
where difference takes the form:
\begin{equation}
    \begin{split}
        \Delta_k &= \bra{\psi_{k}^{gs}}\hat{R}_k\hat{G}_k\hat{R}_k\ket{\psi_{k}^{gs}}\\
        &=\sum_{l\neq gs}\frac{|\bra{\psi_{k}^{gs}}\hat{R}_{k}\ket{\psi_{k}^{l}}|^2}{E_{k}^{gs}-E_{k}^{l}}\le 0.
    \end{split}
    \label{nlf_gs2}
\end{equation}
Here, $E_{k}^{l}>E_{k}^{gs}$ because $E_{k}^{l}$ denotes the energy for eigenstates other than the ground state. Our iteration converges when the equality in Eq. \ref{nlf_gs2} is established. This happens when $\ket{\psi_k^{gs}}$ is an eigenstate of $\hat{R}_k$, consistent with the MLE condition $\hat{R}\ket{\psi}=\ket{\psi}$ (or $\hat{R}\hat{\rho}=\hat{\rho}$). 

Finally, combining Eq. \ref{nlf_gs} and Eq. \ref{nlf_gs2}, we have shown that $M(\ket{\psi_{k+1}^{gs}})-M(\ket{\psi_{k}^{gs}})$ is a negative semi-definite quantity, which proves the theorem. 

One potential complication to the proof is that Eq. \ref{eq:itergs_pertb} needs to assume there is no ground-state level crossing or degeneracy after adding the quantum likelihood gradient. A potential remedy is to keep some low-lying excited states together with the ground state and compare them for maximum likelihood, especially for steps with singular behaviors. Otherwise, we can only hope such transitions are sparse, especially near convergence, and they establish a new line of iterations heading toward the same convergence. A more detailed discussion is available in Ref. \onlinecite{wjb}.

\subsection{Example: $c=\frac{3}{2}$ CFT ground state of Majorana fermion chain}

Here, we consider the spinless 1D Majorana fermion chain model of length $2L$ as an example \cite{PhysRevB.92.235123}:
\begin{equation}
\hat{H}_s=\sum_{j}it\hat{\gamma}_{j}\hat{\gamma}_{j+1}+g\hat{\gamma}_{j}\hat{\gamma}_{j+1}\hat{\gamma}_{j+2}\hat{\gamma}_{j+3},
\label{major}
\end{equation}
where $\hat{\gamma}_{j}$ is the Majorana fermion operator obeying:
\begin{equation}
    \hat{\gamma}_{j}^{\dagger}=\hat{\gamma}_{j}, \{\hat{\gamma}_{i}, \hat{\gamma}_{j}\}=\delta_{ij},
\end{equation}
and $t$ and $g=-1$ are model parameters. This model presents a wealth of nontrivial quantum phases under different $t/g$. We focus on the model parameters in $t/g\in(-2.86, -0.28)$, where the ground state of Eq. \ref{major} is a $c=\frac{3}{2}$ CFT composed of a critical Ising theory ($c=\frac{1}{2}$) and a Luttinger liquid ($c=1$). 

\begin{figure}
    \centering
    \includegraphics[width = 1\linewidth]{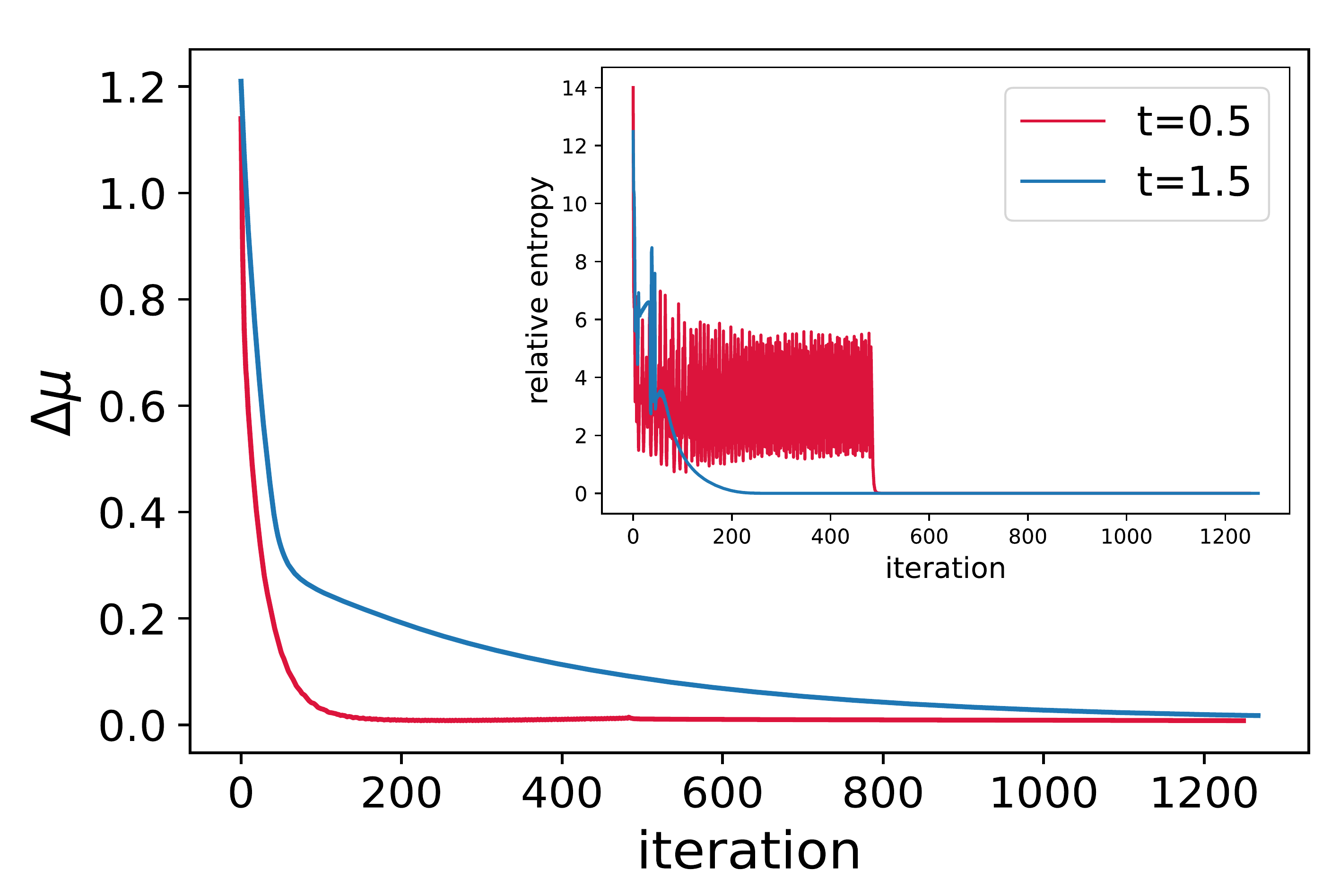}
    \caption{Both the Hamiltonian distance $\Delta\mu$ and the relative entropy $M(\ket{\psi_{k+1}^{gs}})-M_0$ (inset) as defined in the main text indicate successful convergence of the iterations during MLE Hamiltonian learning. We set $g=-1$, $t=0.5$ (red curve) or $t=1.5$ (blue curve), and system size $L=12$ for the target quantum system, and learning rate $\gamma=0.1$ ($\gamma=0.05$) before (after) the 490th iteration. }
    \label{fig:majorana}
\end{figure}

Through the definition of the complex fermions followed by the Jordan-Wigner transformation:
\begin{eqnarray}
    \hat{c}_{j}&=&\frac{\hat{\gamma}_{2j}+i\hat{\gamma}_{2j+1}}{2}, \nonumber\\
    \hat{\sigma}_{j}^{z}&=&2\hat{n}_{j}-1,  \\
    \hat{\sigma}_{j}^{+}&=&e^{-i\pi\sum_{i<j}\hat{n}_{i}}\hat{c}_{j}^{\dagger}, \nonumber
\end{eqnarray}
where $\hat{n}_{j}=\hat{c}_{j}^{\dagger}\hat{c}_{j}$ is the complex fermion number operator, we map Eq. \ref{major} to a 3-local spin chain of length $L$:
\begin{equation}
\begin{split}
   \hat{H}_s=&t\sum_{j}\hat{\sigma}_{j}^{z}-t\sum_{j}\hat{\sigma}_{i}^{x}\hat{\sigma}_{i+1}^{x} \\
   -&g\sum_{j}\hat{\sigma}_{i}^{z}\hat{\sigma}_{i+1}^{z}-g\sum_{j}\hat{\sigma}_{i}^{x}\hat{\sigma}_{i+2}^{x}.\\
\end{split}
\end{equation}

We employ quantum measurements on the ground state $\ket{\psi_s}$ of this Hamiltonian, based on which we carry out our MLE Hamiltonian learning protocol. Here, we evaluate the ground-state properties via exact diagonalization. The numerical results for two cases of $t=0.5, 1.5$ are in Fig. \ref{fig:majorana}. We achieve successful convergence and satisfactory accuracy on the target Hamiltonian. The relative entropy's instabilities are mainly due to the ground state's level crossing and degeneracy.

\subsection{Example: alternative Hamiltonian for ground state}

We have seen that MLE Hamiltonian learning can retrieve the unknown target Hamiltonians via quantum measurements of its Gibbs states, even its ground states. For pure states, however, one interesting byproduct is that the relation between Hamiltonian and eigenstates is essentially many-to-one. Therefore, it is possible to obtain various candidate Hamiltonians $\hat{H}_k$ sharing the same ground state as the original target $\hat{H}_s$, especially by controlling the operator/observable set. Here, we show such numerical examples. 

\begin{figure}
\centering \includegraphics[width = 1\linewidth]{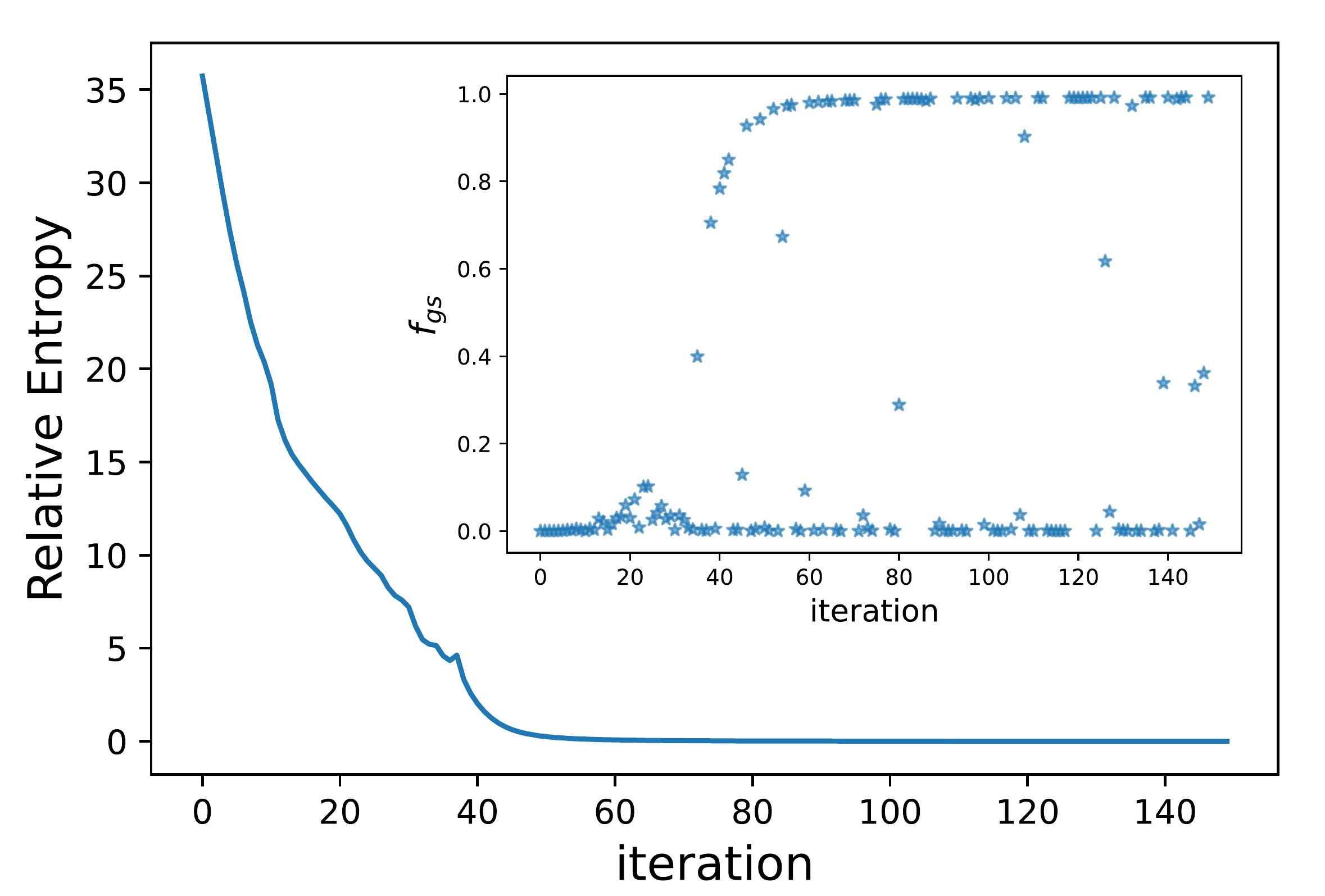}
\caption{Both the relative entropy and the fidelity $f_{gs}=\braket{\psi_s | \psi_{k}^{gs}}$ (inset) indicate successful convergence of the iterations during MLE Hamiltonian learning, yielding a consistent yet different Hamiltonian from the original quantum system. We set the system size $L=15$ and learning rate $\gamma=0.005$.}\label{XZ}
\end{figure}

As our target quantum system, we consider the transverse field Ising model (TFIM) of length $L=15$:
\begin{equation}
\hat{H}_s=J\sum_{j}\hat{S}_{j}^{z}\hat{S}_{j+1}^{z}+g\sum_{j}\hat{S}_{j}^{x},
\end{equation}
at its critical point $J=g=1$. Its ground state is $\ket{\psi_s}$. However, instead of the operators presenting in $\hat{H}_s$, we employ a different operator set for $\ket{\psi_s}$'s quantum measurements:
\begin{equation}
    \{\hat{O}_{i}\} = \{\hat{S}_{i}^{z}\hat{S}_{i+1}^{z}, \hat{S}_{i}^{x}\hat{S}_{i+1}^{x}\}.
    \label{eq:op_set}
\end{equation}
We evaluate the ground-state properties via DMRG.

The subsequent MLE Hamiltonian learning results are in Fig. \ref{XZ}. Since we obtain a candidate Hamiltonian with the operators in Eq. \ref{eq:op_set} and destined to differ from $\hat{H}_s$, the Hamiltonian distance is no longer a viable measure of its accuracy. Instead, we introduce the ground-state fidelity $f_{gs}=\braket{\psi_s|\psi_{k}^{gs}}$, where $\ket{\psi_s}$ ($\ket{\psi_{k}^{gs}}$) is the ground state of $\hat{H}_s$ ($\hat{H}_k$). Interestingly, while the relative entropy shows full convergence, the fidelity $f_{gs}$ jumps between $\sim 99.5\%$ and $\sim 10^{-3}\%$. This is understandable, as the quantum system is gapless, and the ground and low-lying excited states have similar properties under quantum measurements.

\subsection{Example: two-dimensional topological states}

\begin{figure}[h]
    \centering
    \includegraphics[width = 1.03\linewidth]{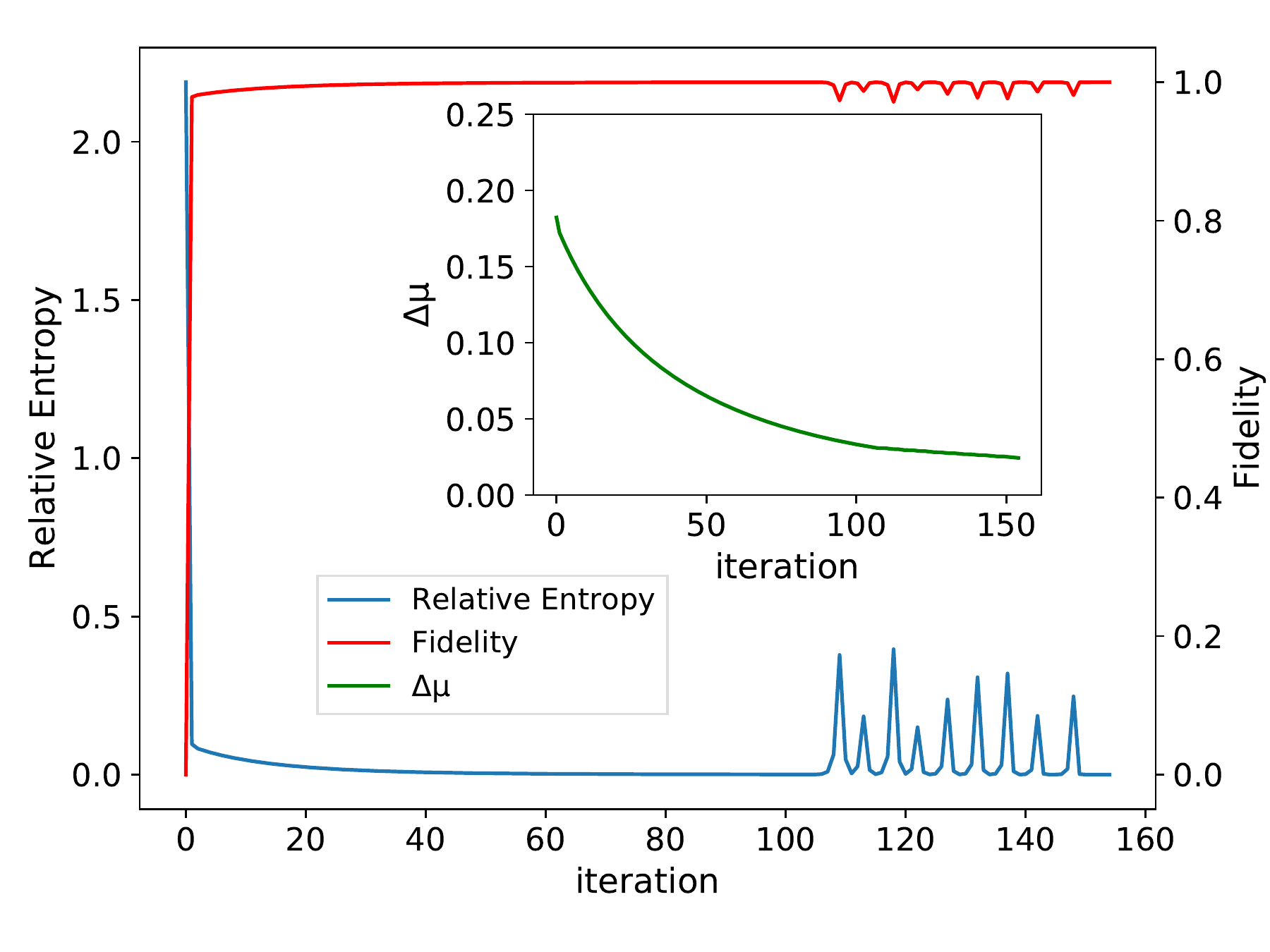}
    \caption{The relative entropy, the fidelity $f_{gs}=\braket{\psi_s | \psi_{k}^{gs}}$, and the Hamiltonian distance $\Delta\mu$ (inset) show distinct convergence behaviors in the iterations of MLE Hamiltonian learning for a 2D topological CSL system. Our system size is $4 \times 4$, and we set the learning rate $\gamma=0.1$.}
    \label{CSLConstruct}
\end{figure}

Here, we consider MLE Hamiltonian learning on two-dimensional topological quantum systems. In particular, we consider the chiral spin liquid (CSL) on a triangular lattice:
\begin{equation}
    \hat{H}_s =J_1\sum_{\langle ij\rangle}\Vec{S}_i\cdot\Vec{S}_j+J_2\sum_{\langle\langle ij\rangle\rangle}\Vec{S}_i\cdot\Vec{S}_j+K\sum_{i,j,k\in \bigtriangledown/\bigtriangleup}\Vec{S}_i\cdot\left(\Vec{S}_j\times\Vec{S}_k\right),
\end{equation}
where the first and second terms are Heisenberg interactions, and the last term is a three-spin chiral interaction. Previous DMRG studies have established $\hat{H}_s$'s ground state as a CSL under the model parameters $J_1 = 1.0$, $J_2=0.1$, and $K=0.2$\cite{PhysRevB.96.075116}, which we set as the parameters of the target Hamiltonian. Here, we employ exact diagonalization on a $4 \times 4$ system. Based upon entanglement studies of the lowest-energy eigenstates, we verify that both the modular $SU$ matrix corresponding to $C_6$ rotations and the entanglement entropy fit well with a CSL topological phase\cite{PhysRevB.85.235151}. Subsequently, we perform MLE Hamiltonian learning based on quantum measurements of the ground state, focusing on the operators presenting in $\hat{H}_s$. We summarize the results in Fig. \ref{CSLConstruct}. The Hamiltonian distance indicates a stable converging accuracy, yet the relative entropy and the fidelity $f_{gs}=\braket{\psi_s | \psi_{k}^{gs}}$ witness certain instabilities. Indeed, being a topological phase means ground-state degeneracy - competing low-energy eigenstates with global distinctions yet similar local properties. 

\end{document}